\documentstyle[12pt,epsfig]{article}
\newcounter{popnr}

\def\theequation{\thesection.\arabic{equation}}
\renewcommand{\theequation}{\arabic{section}.\arabic{equation}}

\def\be{\begin{equation}}
\def\ee{\end{equation}}
\def\lsim{\raise0.3ex\hbox{$<$\kern-0.75em\raise-1.1ex\hbox{$\sim$}}}
\def\gsim{\raise0.3ex\hbox{$>$\kern-0.75em\raise-1.1ex\hbox{$\sim$}}}

\begin{document}

\title{
{\normalsize
\vspace*{-1cm}
%\hfill hep-ph/xxxxxxx\\ 
\hfill BI-TP 2001/01 \\ \vspace*{1cm}
}
{\bf The generalized vector dominance/colour-dipole
picture of deep-inelastic scattering at low x}\thanks{Supported by the
BMBF, Bonn, Germany, Contract 05 HT9PBA2}}

\author{\and
{\bf G. Cveti\v c} \\
Dept.~of Physics, Universidad T\'ecnica Federico Santa Mar\'{\i}a,\\
Valpara\'{\i}so, Chile \\[1.2mm]
\and
{\bf D. Schildknecht} \\
Fakult\"{a}t f\"{u}r Physik, Universit\"{a}t Bielefeld\\
D-33501 Bielefeld \\[1.2mm]
\and
{\bf  B. Surrow} \\
%Max-Planck Institut f\"{u}r Physik (Werner-Heisenberg-Institut) \\
%F\"{o}hringer Ring 6, D-80805 M\"{u}nchen
~~~~~~~~~~~~~~~~~~~~~~ DESY, D-22607 Hamburg ~~~~~~~~~~~~~~~~~~~~~~~~~~
\\[1.2mm]
\and
{\bf M. Tentyukov}\thanks{On leave from BLTP JINR, Dubna, Russia} \\
Fakult\"{a}t f\"{u}r Physik, Universit\"{a}t Bielefeld \\[1.2mm]
D-33501 Bielefeld
}
\date{}
\maketitle
\thispagestyle{empty}

{\abstract
We give a detailed account of the recently formulated generalized vector
dominan\-ce/co\-lour-dipole picture (GVD/CDP) of deep-inelastic scattering
at low $x\cong Q^2/W^2$,
including photoproduction. The approach, based on $\gamma^*
(q \bar q)$ transitions, $q \bar q$ propagation and diffractive $(q \bar q)
p$ scattering via the generic structure of the two-gluon exchange,
provides a unique and quantitatively successful theory for the $\gamma^* p$
total cross
section, $\sigma_{\gamma^* p} (W^2,Q^2)$, at low $x$.
The GVD/CDP is shown to imply the
empirical low-$x$ scaling law,
$\sigma_{\gamma^* p} (W^2,Q^2)=\sigma_{\gamma^* p} (\eta)$ with
$\eta=(Q^2+m_0^2)/\Lambda^2(W^2)$, that was established by a
model-independent analysis of the experimental data.

}

\newpage
\addtocounter{page}{-1}

\section{Introduction}
Two important observations \cite{HERA} on deep inelastic scattering
(DIS) at low values of the Bjorken scaling variable,
$x_{bj} \cong Q^2/W^2 \ll 1$, were made since HERA
started running in 1992:

\medskip
i)
Diffractive production of high-mass states of masses
$M_X \lsim 30$GeV
at an appreciable rate relative to the total virtual-photon-proton cross
section, $\sigma_{\gamma^* p} (W^2 , Q^2)$.
The sphericity and thrust analysis of the diffractively produced states
revealed \cite{HERA} (approximate) agreement in shape with the final state
produced in
$e^+ e^-$ annihilation at the energy $\sqrt s = M_X$. This observation of high-mass
diffraction confirmed  the conceptual basis of generalized
vector dominance (GVD) \cite{SaSchi}
that generalizes the role of the low-lying vector mesons in
photoproduction \cite{Stod} to
DIS at low $x$ via the inclusion of high-mass
contributions\footnote{Indirect empirical evidence for diffractive
production of high-mass states was previously provided by the observation
\cite{achman} of
shadowing in DIS on complex nuclei at low $x$ and large $Q^2$ \cite{sch73}.
Compare also \cite{Stodolsky} for the connection between shadowing and
high-mass diffractive production.}.

\medskip
ii)
An increase of $\sigma_{\gamma^* p} (W^2, Q^2)$ with increasing energy
at fixed $Q^2$ and low $x$ considerably stronger
than the smooth ``soft-pomeron'' behaviour known from photoproduction
and hadron-hadron scattering.

In a brief communication \cite{Surrow}, we recently reported the empirical
validity of a scaling law for the $Q^2$ dependence and the $W$ dependence of
the virtual-photon-proton cross
section,
\begin{equation}\label{1.1}
         \sigma_{\gamma^* p} (W^2 , Q^2)=\sigma_{\gamma^* p} \Big(\eta\Big),
\end{equation}
and, moreover, we noted that the scaling law (\ref{1.1}) follows from the
generalized vector dominan\-ce/co\-lour-dipole
picture (GVD/CDP).
This picture of DIS at low $x$ rests \cite{Cvetic} on $\gamma^*(q\bar q)$
transitions, propagation of the $q\bar q$ state and its forward scattering
from the proton via the generic structure of two-gluon exchange
\cite{a}\footnote{
The ``generic structure'' of two-gluon exchange includes exchange of more
than two gluons, exchange of a gluon ladder, etc. The only relevant and
essential point is equal strength and opposite sign of the two generic
diagrams depicted in fig.\ref{fig1}.
}.
Accordingly, the GVD/CDP supplements the traditional (off-diagonal) GVD
approach \cite{SaSchi,FRS} by taking into account the $q\bar q$
configuration in the $\gamma^*(q\bar q)$ transition, as well as the generic
structure of the two-gluon-exchange interaction, the $q\bar q$ colour dipole
is subject to \cite{cetrar5}, when transversing the proton. The dimensionless
low-$x$ scaling variable $\eta$ in (\ref{1.1}) is given by
$\eta (\Lambda^2(W^2),Q^2)=(Q^2+m_0^2)/\Lambda^2(W^2)$,
where $\Lambda^2(W^2)$ is an increasing
function of  $W^2$ and $m_0^2$ denotes a threshold mass.

In the present
paper, we provide a detailed account of our recent findings.
In Section 2, we will give an explicit analytic representation
of $\sigma_{\gamma^* p} (W^2 , Q^2)$ in the GVD/CDP. We will derive the
scaling law (\ref{1.1}), and we will discuss the photoproduction limit and the
relation of the present approach to the pre-QCD formulation of off-diagonal
GVD. In Section 3, we will present the model-independent analysis of the
experimental data that establishes the empirical validity of the scaling law
(\ref{1.1}).
Subsequently, we will
show that the observed $\eta$ dependence of the data coincides with the
one predicted by the
GVD/CDP. Final conclusions will be given in Section 4.

\section{The generalized vector dominance/colour-dipole picture (GVD/CDP).}
\setcounter{equation}{0}
\subsection{Generalities}

We follow custom \cite{Forshaw,Golec} and as a starting point
 adopt a representation of
the virtual-photon-proton cross section, $\sigma_{\gamma^*_{T,L}p} (W^2,
Q^2)$, in transverse position space. Subsequently, we transform to momentum
space, rather than proceeding in historical
order \cite{cetrar5,Cvetic} from
momentum space to transverse position space.

In transverse position space, accordingly, we represent $\sigma_{\gamma^*_{T,L}
p} (W^2,Q^2)$ as an integral over the variables $\vec r_\perp$ and $z$
determining the $(q \bar q)$ configuration in the $\gamma^* (q \bar q)$
transition \cite{Cvetic}\footnote{
By definition, $\vec r_\perp$ denotes the transverse (two-dimensional)
vector of the quark-antiquark separation. The (light-cone) variable $z$ is
related to the angle of the quark momentum in the rest frame of the $q\bar q$
system via \cite{Cvetic} $4z(1-z)=\sin^2\theta$, the resulting mass of the
$q\bar q$ state thus being given by (\ref{2.6}) below.
Twice the helicity of the quark and antiquark is denoted by $\lambda$ and
$\lambda^\prime$.
},

\be
\sigma_{\gamma^*_{T,L}p} (W^2,Q^2) = \sum_{\lambda, \lambda^\prime = \pm 1}
\int dz \int d^2r_\perp \vert \psi^{(\lambda, \lambda^\prime)}_{T,L}
(\vec r_\perp, z, Q^2)\vert^2 \sigma_{(q \bar q)p}
(\vec r^{~2}_\perp, z,W^2).
\label{2.1}
\ee

\begin{figure}[ht]
\begin{minipage}[b]{.49\linewidth}
 \centering\epsfig{file=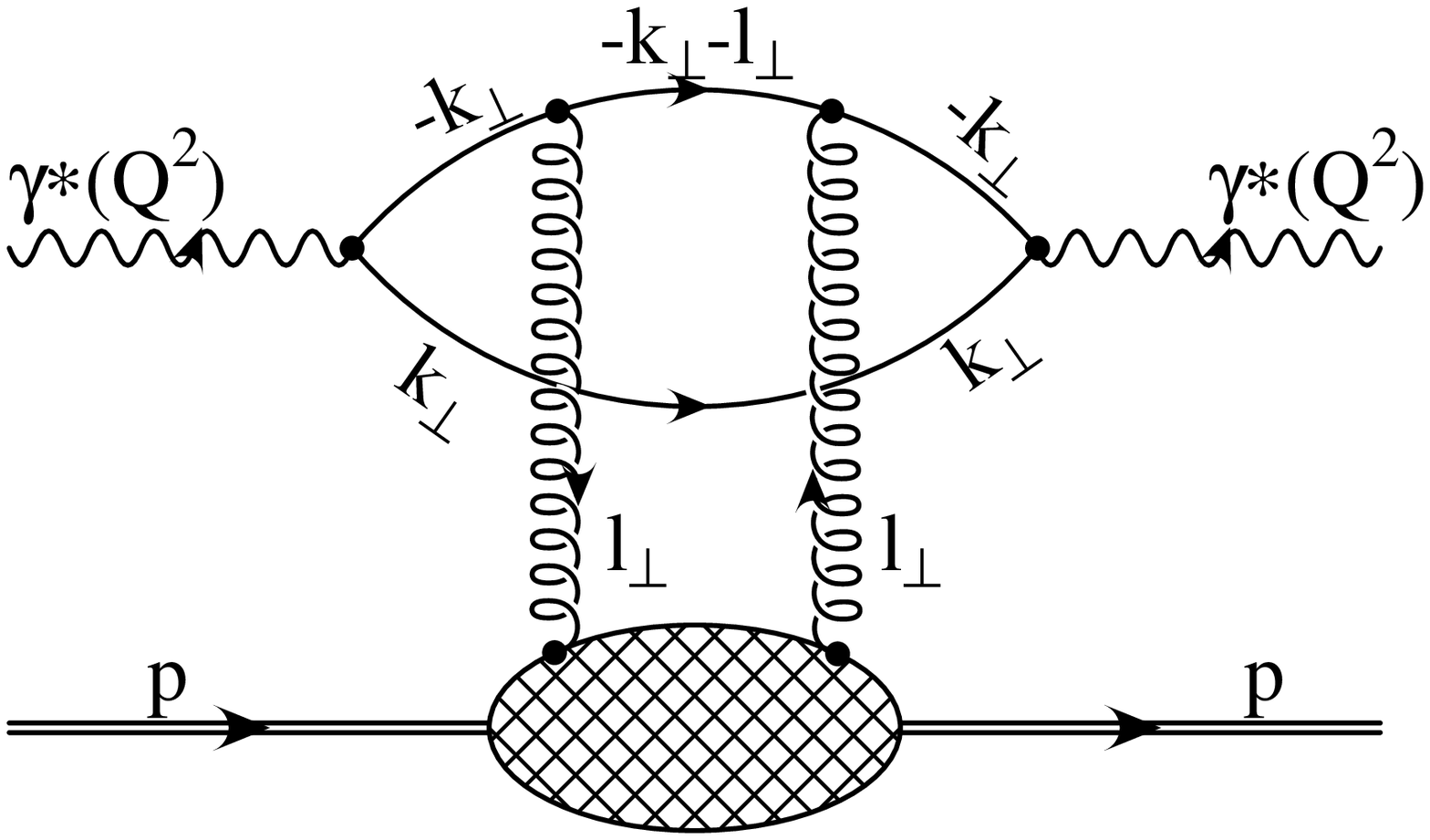,width=\linewidth}
\end{minipage}
\begin{minipage}[b]{.49\linewidth}
 \centering\epsfig{file=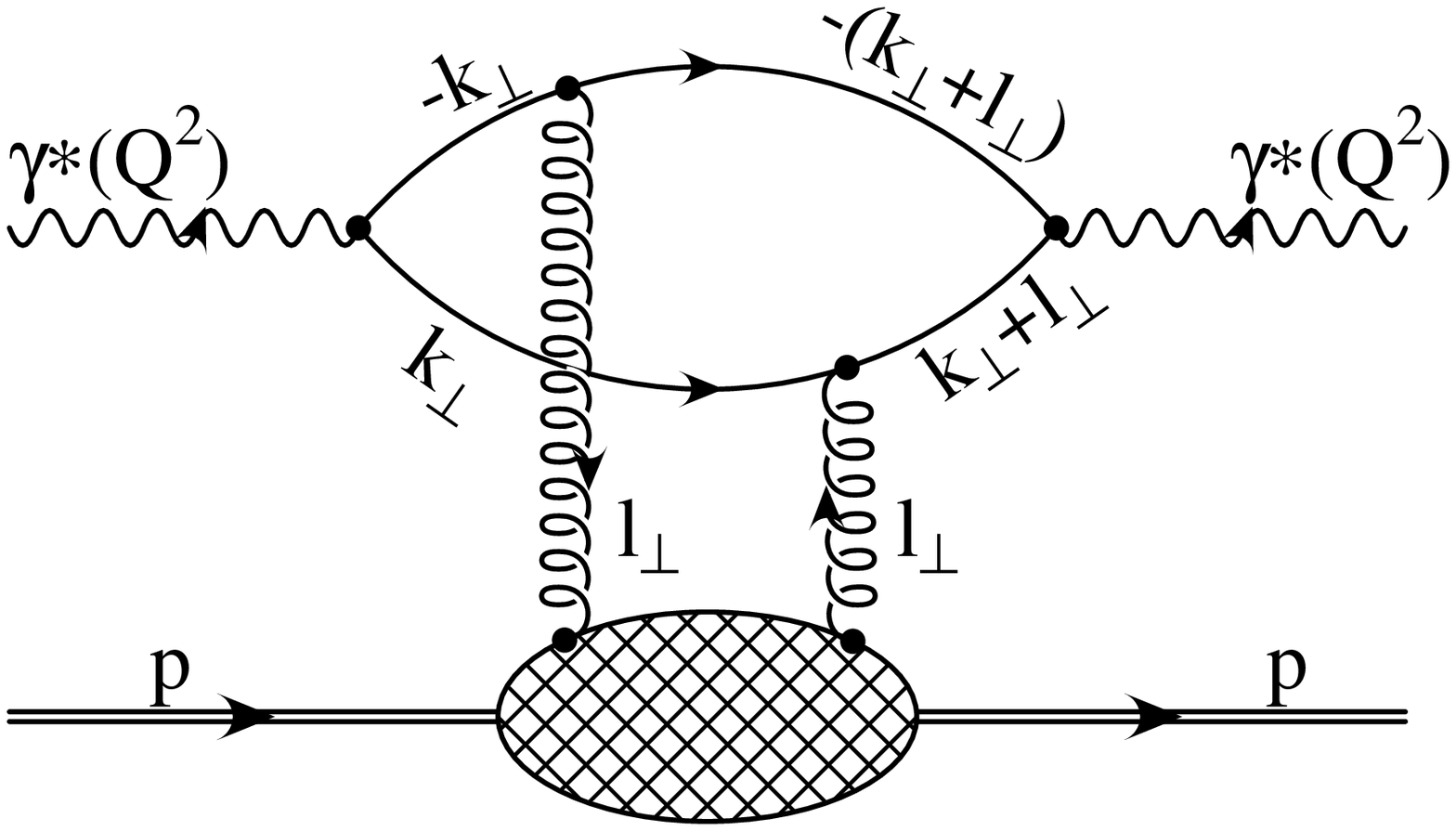,width=\linewidth}
\end{minipage}
\vspace{-1.0cm}
\caption{The two-gluon exchange. 
The arrows relate to the transverse-momentum flow.
   \label{fig1}}
\end{figure}

\noindent
The $\gamma^* (q \bar q)$ transition amplitude, known as the photon wave
function, for transverse $(T)$ and longitudinal $(L)$ photons is described by
$\psi^{(\lambda, \lambda^\prime)}_{T,L} (r_\perp, z, Q^2)$.
Implicitly, (\ref{2.1}) contains the assumption that the configuration variable
$z$ remains unchanged during the $(q \bar q)p$ scattering process. The
representation (\ref{2.1}) {\em must} be read in conjunction with \cite{Cvetic}

\be
\sigma_{(q \bar q)p} (\vec r^{~2}_\perp, z, W^2) = \int d^2 l_\perp \tilde \sigma_{(q
\bar q)p} (\vec l^{~2}_\perp, z, W^2)
(1 - e^{-i \vec l_\perp \cdot \vec r_\perp}),
\label{2.2}
\ee

\noindent
that implies

\be
\sigma_{(q \bar q)p} (\vec r^{~2}_\perp, z,W^2) \to \left\{ \begin{array}{c@{\quad,
\quad}l}
\vec r^{~2}_\perp\cdot \frac{\pi}{4}
\int d\vec l^{~2}_\perp \cdot \vec l^{~2}_\perp
\ \tilde \sigma_{(q \bar q)p} (\vec l^{~2}_\perp, z,W^2)
 & {\rm for}~ r_\perp \to 0,\\
\int d^2l_\perp \tilde \sigma_{(q \bar q)p} (\vec l^{~2}_\perp, z,W^2) & {\rm for}~
r_\perp \to \infty.
\end{array} \right.
\label{2.3}
\ee
The two-dimensional vector $\vec l_\perp$ is to be identified with the transverse
(gluon) momentum absorbed or emitted by the quark (compare fig.\ref{fig1}).
The vanishing of the colour-dipole cross section, $\sigma_{(q \bar q)p} (\vec
r^{~2}_\perp, z, W^2)$,
for vanishing transverse interquark separation is known as
``colour transparency'' \cite{cetrar5}.

The generic two-gluon-exchange structure (compare fig.\ref{fig1})
contained in (\ref{2.2}) becomes
explicit when inserting (\ref{2.2}) into (\ref{2.1}) in conjunction with the
Fourier
representation of the photon wave function,

\be
\psi^{(\lambda,\lambda^\prime)}_{T,L} (\vec r_\perp, z, Q^2) =
{{\sqrt{4 \pi}} \over {16 \pi^3}} \int_{k_{\perp 0}} d^2 k_\perp exp (i \vec
k_\perp \cdot \vec r_\perp) {\cal M}^{(\lambda, \lambda^\prime)}_{T,L}
(\vec k_\perp, z; Q^2).
\label{2.4}
\ee

\noindent
One obtains (cf. \cite{Cvetic})
\pagebreak
\begin{eqnarray}\label{2.5}
\lefteqn{\sigma_{\gamma^*_{T,L}p}
 (W^2,Q^2)}\nonumber \\
 & & = {N_c \over {16 \pi^3}} \sum_{\lambda,
\lambda^\prime = \pm 1} \int dz \int d^2 l_\perp \tilde \sigma_{(q \bar q)p}
(\vec l^{~2}_\perp, z, W^2)\nonumber \\
\times & &\hspace*{-0.5cm}
\int_{\vert \vec k_\perp \vert \geq k_{\perp 0}}
d^2 k_\perp \int_{\vert \vec k^{~\prime}_\perp \vert \geq k_{\perp 0}}
d^2k^\prime_\perp \nonumber \\
\times & &\hspace*{-0.5cm} {\cal M}^{(\lambda, \lambda^\prime)}_{T,L}
(\vec k^\prime_\perp, z; Q^2)^*
{\cal M}^{(\lambda, \lambda^\prime)}_{T,L} (\vec k_\perp, z; Q^2)
\nonumber \\
\times & &\hspace*{-0.5cm} \left[ \delta (\vec k^\prime_\perp - \vec k_\perp)
- \delta
(\vec k^\prime_\perp - \vec k_\perp - \vec l_\perp) \right],
\end{eqnarray}

\noindent
where $N_c=3$ denotes the number of quark colours.
The amplitudes ${\cal M}_{T,L}^{(\lambda, \lambda^\prime)} (\vec k_\perp, z;
Q^2)$ in (\ref{2.4}) and (\ref{2.5}) contain \cite{Cvetic}
the couplings of the photon to
the $q \bar q$
pair as well as the propagators of the $q \bar q$ pair of mass
$M_{q \bar q}$, where in terms of the quark (antiquark) transverse momentum,
$\vert \vec k_\perp\vert$,
\be
M^2_{q \bar q} = {{\vec k^{~2}_\perp} \over {z(1-z)}}.
\label{2.6}
\ee

\noindent
Transitions diagonal and off-diagonal in the masses of the initial and final
$q \bar q$ states, $M_{q \bar q}$ from (\ref{2.6}), and $M^\prime_{q \bar q}$
according to

\be
M^{\prime 2}_{q \bar q} = {{(\vec k_\perp + \vec l_\perp)^2} \over
{z(1-z)}},
\label{2.7}
\ee

\noindent
contribute with equal weight to (\ref{2.5}), but opposite in sign, as
required by
the generic two-gluon-exchange structure.

Conversely, it is precisely the
generic two-gluon-exchange structure of the
forward-virtual-Compton-scattering
amplitude that justifies (\ref{2.1}) as a starting point for
low-x DIS.

According to (\ref{2.3}), $\tilde \sigma_{(q \bar q)p}
(\vec l^{~2}_\perp, z, W^2)$
should
vanish sufficiently rapidly to yield a convergent integral. It
may be suggestive to assume a Gaussian in $\vec l^{~2}_\perp$ for $\tilde
\sigma_{(q \bar q)p} (\vec l^{~2}_\perp, z, W^2).$
Actually, explicit calculations become much simpler if, without much loss of
generality,
instead of a Gaussian a $\delta$-function, located at a finite value of
$\vec l^{~2}_\perp$,
is used as an effective description of
$\tilde{\sigma}_{(q\bar{q}p)}(\vec l^{~2}_\perp,z,W^2)$.

Accordingly, we adopt the simple ansatz\cite{Surrow},

\be
\tilde{\sigma}_{(q\bar{q}p)}(\vec l^{~2}_\perp,z,W^2) = \sigma^{(\infty)} (W^2)
{1 \over \pi} \delta (\vec l^{~2}_\perp - z (1-z) \Lambda^2 (W^2)).
\label{2.8}
\ee

\noindent
This ansatz associates with any given energy, $W$, an (effective) fixed value
of the two-dimensional (gluon) momentum transfer, $\vert \vec l_\perp \vert$,
determined by the so far unspecified function $\Lambda (W^2)$. The ansatz
(\ref{2.8}) also incorporates the assumption that `aligned', $z \to 0$,
configurations\cite{Bjorken} of the $q \bar q$ pair absorb vanishing,
$\vec l^{~2}_\perp \to 0$, gluon momentum.

For the subsequent interpretation of
our results, we note the explicit form of the transverse-position-space
colour-dipole cross section, obtained by substituting (\ref{2.8}) into
(\ref{2.2}),

\begin{eqnarray}\label{2.9}
& & \sigma_{(q \bar q)p} (\vec r^{~2}_\perp, z, W^2) = \sigma^{(\infty)} (W^2)
(1 - J_0 (r_\perp \cdot \sqrt{z(1-z)} \Lambda (W^2)))\nonumber \\[1mm]
\simeq & & \sigma^{(\infty)} (W^2) \cdot \left\{
\begin{array}{l@{\quad~\quad}l}
{1 \over 4} z (1-z) \Lambda^2 (W^2) \vec r^{~2}_\perp, & {\rm for}~ {1 \over 4} z
(1-z) \Lambda^2 (W^2) \vec r^{~2}_\perp \to 0,\\[3mm]
1, & {\rm for}~{1 \over 4} z (1-z) \Lambda^2 (W^2) \vec r^{~2}_\perp \to \infty.
\end{array} \right.
\end{eqnarray}

\noindent
The limit of $\sigma^{(\infty)}(W^2)$ in the second line of the
approximate equality in (\ref{2.9}) stands for an oscillating
behaviour with decreasing amplitudes
of the Bessel function, $J_0 (r_\perp
\sqrt{z(1-z)} \Lambda (W^2))$, around $\sigma^{(\infty)}(W^2)$, when its
argument tends towards infinity. Apart from these oscillations, the
behaviour of $\sigma_{(q \bar q)p} (\vec r^{~2}_\perp, z, W^2)$ in (\ref{2.9}) is
identical to the one obtained, if the $\delta$-function in (\ref{2.8}) were
replaced by a Gaussian. Concerning the high-energy behaviour of
$\sigma_{(q \bar q)p} (\vec r^{~2}_\perp, z, W^2)$, we note that it is consistent
with unitarity restrictions, provided a decent high-energy behaviour is
imposed on $\sigma^{(\infty)} (W^2)$.

The ansatz (\ref{2.8}) is to be seen as an effective realization, without
much loss of generality,
of the underlying requirements of colour transparency, (\ref{2.2}), (\ref{2.3}),
and hadronic unitarity for the colour-dipole cross section. The unitarity
requirement enters via the aforementioned
decent high-energy behaviour of $\sigma^{(\infty)}
(W^2)$.

With (\ref{2.8}), the virtual-photon-proton cross section (2.5)
may be simplified
considerably (cf. \cite{Cvetic} as well as Appendix A).
The right-hand side becomes reduced to essentially the
product of $\sigma^{(\infty)} (W^2)$ from (\ref{2.8}) with a dimensionless
integral
over the masses $dM^2 \equiv dM^2_{q \bar q}$ and $dM^{\prime 2} \equiv
dM^{\prime 2}_{q \bar q}$ from (\ref{2.6}) and (\ref{2.7}). The dimensionless
integral depends on the ratios of the available parameters, namely
$Q^2/\Lambda^2 (W^2)$ and $m^2_0/\Lambda^2 (W^2)$, where $m^2_0$ stems
from the lower limit of the integral in (\ref{2.5}), where $k^2_{\perp 0} =
z (1-z) m^2_0$. This threshold mass corresponds to the fact that the
masses of hadronic vector states lie above a flavour-dependent lower limit.
For non-strange quarks, we have $m^2_0 \lsim m^2_\rho$, i.e. $m^2_0$ is
identified as the mass scale at which $e^+e^- \to~ {\rm hadrons}$ reaches
appreciable strength.

Instead of $Q^2/\Lambda^2 (W^2)$ and $m^2_0/\Lambda^2
(W^2)$, it will turn out to be preferable to use the low-x scaling
variable\cite{Surrow}

\be
\eta\Big(\Lambda^2(W^2),Q^2\Big) = {{Q^2 + m^2_0} \over {\Lambda^2 (W^2)}}
\label{2.10}
\ee
in conjunction with $m^2_0/\Lambda^2(W^2)$. The virtual-photon-proton
cross section (\ref{2.5}) then becomes (compare Appendix A)

\be
\sigma_{\gamma^*_{T,L}p} (W^2, Q^2) = {{\alpha R_{e^+e^-}} \over
{3 \pi}} \sigma^{(\infty)} (W^2) I_{T,L} (\eta, {{m^2_0} \over
{\Lambda^2(W^2)}}).
\label{2.11}
\ee
The quark charges $Q_i$ in units of the positron charge enter (\ref{2.11}) via

\be
R_{e^+e^-} = 3 \sum Q^2_i,
\label{2.12}
\ee

\noindent
that is the ratio of hadron production to $\mu^+\mu^-$ pair production
in $e^+ e^-$ annihilation. When specifying (\ref{2.11}) to photoproduction,
$Q^2=0$, only the light quark flavours ($u,\ d,\ s$) contribute
appreciably, and, accordingly, $R_{e^+e^-}=2$ is to be inserted.

The function $I_{T,L} (\eta, m^2_0/\Lambda^2(W^2))$ in (\ref{2.11}) is
conveniently
split into two additive contributions, a dominant term, $I^{(1)}_{T,L}$,
and a correction term, $I_{T,L}^{(2)}$. This splitting will allow us to
derive exact analytical expressions for one of the terms, the dominant one,
while for the
correction term, we will be content with an approximation in analytical form.

The integral representations for the transverse and longitudinal dominant
parts, $I^{(1)}_T$ and $I^{(1)}_L$ read,

\begin{eqnarray}\label{2.13}
I^{(1)}_T \left( \eta, {{m^2_0} \over
{\Lambda^2(W^2)}}\right) & &\hspace*{-0.5cm} =
{1 \over \pi} \int^\infty_{m^2_0} dM^2 \int^{(M+\Lambda (W^2))^2}_{(M-\Lambda
(W^2))^2} dM^{\prime 2}\omega (M^2, M^{\prime 2}, \Lambda^2 (W^2))\nonumber \\
& &\hspace*{-1.5cm}
\times
\left[
   \frac{M^2}{(Q^2+M^2)^2}-
   \frac{M^{\prime 2}+M^2-\Lambda^2 (W^2)}
        {2(Q^2+M^2)(Q^2+M^{\prime 2})}
\right],
\end{eqnarray}

and

\begin{eqnarray}\label{2.14}
I^{(1)}_L  \left( \eta, {{m^2_0} \over
{\Lambda^2(W^2)}}\right) & &\hspace*{-0.5cm}=
{1 \over \pi} \int^\infty_{m^2_0} dM^2 \int^{(M+\Lambda (W^2))^2}_{(M-\Lambda
(W^2))^2} dM^{\prime 2}\nonumber \omega (M^2, M^{\prime 2}, \Lambda^2 (W^2))\\
& &\hspace*{-0.5cm}
\times
\left[
   \frac{Q^2}{(Q^2+M^2)^2}-
   \frac{Q^2}
        {(Q^2+M^2)(Q^2+M^{\prime 2})}
\right].\end{eqnarray}

\noindent
The correction terms take the form

\begin{eqnarray}\label{2.15}
I^{(2)}_T \left( \eta, {{m^2_0} \over
{\Lambda^2 (W^2)}} \right) & &\hspace*{-0.5cm} =
{1 \over \pi} \int^{\infty}_{m^2_0} dM^2 \Theta\left(m^2_0-(M-\Lambda(W^2))^2\right)
\int^{m^2_0}_{(M-\Lambda (W^2))^2} dM^{\prime 2} \nonumber \\
& &\hspace*{-0.5cm}  \times \omega (M^2, M^{\prime 2},
\Lambda^2 (W^2))
{{M^{\prime 2} + M^2 - \Lambda^2 (W^2)} \over
{2 (Q^2 + M^2) (Q^2 + M^{\prime 2})}},
\end{eqnarray}

\noindent
and
\begin{eqnarray}\label{2.16}
I^{(2)}_L  \left( \eta, {{m^2_0} \over
{\Lambda^2 (W^2)}} \right) & &\hspace*{-0.5cm} =
{1 \over \pi} \int^{\infty}_{m^2_0} dM^2 \Theta\left(m^2_0-(M-\Lambda(W^2))^2\right)
\int^{m^2_0}_{(M-\Lambda (W^2))^2} dM^{\prime 2}\nonumber \\
& &\hspace*{-0.5cm} \times \omega (M^2, M^{\prime 2},
\Lambda^2 (W^2)) {{Q^2} \over
{(Q^2 + M^2) (Q^2 + M^{\prime 2})}}.
\end{eqnarray}

\noindent
Replacing the $\Theta$ function in (\ref{2.15}) and in (\ref{2.16}) by
the integration limits in the integration over $dM^2$, we
have
\begin{eqnarray}\label{2.16a}
\lefteqn{ \int^{\infty}_{m^2_0} dM^2
 \Theta\left(m^2_0-(M-\Lambda(W^2))^2\right)\int^{m^2_0}_{(M-\Lambda
 (W^2))^2} dM^{\prime 2}...}\\[1mm]
&&=\left\{
\begin{array}{ll}
    \int^{(m_0+\Lambda(W^2))^2}_{m_0^2}dM^2\int^{m_0^2}_{(M-\Lambda(W^2))^2}
    dM^{\prime 2}...
    &\mbox{, for } \Lambda(W^2)<2m_0,\\[3mm]
    \int^{(\Lambda(W^2)+m_0)^2}_{(\Lambda(W^2)-m_0)^2}dM^2
    \int^{m_0^2}_{(M-\Lambda(W^2))^2} dM^{\prime 2}...
    &\mbox{, for } \Lambda(W^2)>2m_0,\nonumber
\end{array}
\right.
\end{eqnarray}
i.e.  the terms (\ref{2.15}) and (\ref{2.16}),  when added to the
dominant terms (\ref{2.13}) and (\ref{2.14}), respectively, assure
that the lower limit of the integration  over $dM^{\prime 2}$ is given
by $M^{\prime 2}=m_0^2$,  and coincides with  the lower limit of  the
integration over $dM^2$, as required by symmetry in the incoming and
outgoing $q\bar q$ masses.
Actually it
turns out that the transverse correction term, $I^{(2)}_T$, is negligible,
while the longitudinal one\footnote{This is connected with the relative
enhancement of low masses in the integrand of the longitudinal case versus
the transverse one by the factor of $Q^2$.} is of some importance.

For the explicit expression for the integration measure $\omega (M^2,
M^{\prime 2}, \Lambda^2 (W^2))$ appearing in (\ref{2.13}) to (\ref{2.16})
we refer
to Appendix A. For convenient reference, we note the integral relations
\cite{Cvetic}

\be
{1 \over \pi} \int^{(M + \Lambda (W^2))^2}_{(M - \Lambda (W^2))^2}
dM^{\prime 2} \omega (M^2, M^{\prime 2}, \Lambda^2 (W^2)) = 1,
\label{2.17}
\ee

and

\be
{1 \over \pi} \int^{(M + \Lambda (W^2))^2}_{(M-\Lambda (W^2))^2}
dM^{\prime 2} \omega (M^2, M^{\prime 2}, \Lambda^2 (W^2)) M^{\prime 2}
= M^2 + \Lambda^2 (W^2),
\label{2.18}
\ee
however.

\subsection{Analytic evaluation of $\sigma_{\gamma^* p}(W^2, Q^2)$}

We concentrate on the unpolarized cross section,

\be
\sigma_{\gamma^*p} (W^2, Q^2) = \sigma_{\gamma^*_T p} + \sigma_{\gamma^*_L p},
\label{2.19}
\ee

\noindent
and refer to Appendix B for a separate treatment of the longitudinal and
transverse parts.

In terms of the sums of the transverse and longitudinal contributions in
(\ref{2.13}) and (\ref{2.14}),

\be
I^{(1)} = I^{(1)}_T + I^{(1)}_L,
\label{2.20}
\ee

\noindent
and in (\ref{2.15}) and (\ref{2.16}),

\be
I^{(2)} = I^{(2)}_T + I^{(2)}_L,
\label{2.21}
\ee

\noindent
and upon taking the sum of the dominant and the correction part,

\be
I \left( \eta , {{m^2_0} \over {\Lambda^2(W^2)}} \right) = I^{(1)} \cdot
\left( 1 + {{I^{(2)}} \over {I^{(1)}}} \right),
\label{2.22}
\ee

\noindent
with (\ref{2.11}), the unpolarized cross section (\ref{2.19}) becomes

\be
\sigma_{\gamma^*p} (W^2, Q^2) = {{\alpha R_{e^+e^-}} \over {3 \pi}}
\sigma^{(\infty)} (W^2) I \left( \eta , {{m^2_0} \over {\Lambda^2(W^2)}}
\right) .
\label{2.23}
\ee

As mentioned, the integral representations for the dominant transverse and
longitudinal contributions (\ref{2.13}) and (\ref{2.14}) may be analytically
evaluated in a straight-forward manner. Accordingly, $I^{(1)}$ in (\ref{2.20})
and (\ref{2.22}) is explicitly given by

\begin{eqnarray}\label{2.24}
\nonumber
\lefteqn{
   I^{(1)}
   \left( \eta,\mu \equiv
                 \frac{
                         m_0^2
                      }{
                         \Lambda^2(W^2)
                      }
   \right)
}\\
%&&
\nonumber
= \frac{1}{2} \hspace*{-0.5cm}&&\ln
\frac{
        \eta-1+
        \sqrt{(1+\eta)^2-4\mu}
     }{
        2\eta
     }
 \\
%&&
%\nonumber
+ \frac{
       1
     }{
       2
       \sqrt{1+4(\eta-\mu)}
     }
\times \hspace*{-0.5cm}&&\ln
\frac{
       \eta
       \left(
             1+\sqrt{1+4(\eta-\mu)}
       \right)
     }{
       4\mu-1-3\eta+
       \sqrt{
           \Big(
               1+4(\eta-\mu)
           \Big)
           \Big(
               (1+\eta)^2-4\mu
           \Big)
       }
     }.
\end{eqnarray}

The correction term $I^{(2)}$ in (\ref{2.21}), containing the sum of the
integrals in
(\ref{2.15}) and (\ref{2.16}), was evaluated by numerical integration
for various
sets of values of $\eta$ and $\mu$. Guided by these numerical results, we
found a simple analytic approximation formula for the ratio in
(\ref{2.22}) that reads

\begin{equation}
1+\frac{
         I^{(2)}
       }{
         I^{(1)}
       }
=1+2
\frac{
        m_0^2
     }{
        \Lambda^2
     }
\sqrt{
       \frac{1}{2}+ \frac{1}{\pi}{\rm arctg}
       \left(
           \frac{1}{\pi}
           \left(
              \eta-\frac{m_0^2}{\Lambda^2}
           \right)
           -
           \frac{
                   1
                }{
                   2
                   \left(
                       \eta-\frac{m_0^2}{\Lambda^2}
                   \right)
                }
       \right)
}.
\label{2.25}
\end{equation}

\begin{figure}[ht]
%\vspace*{-1.5cm}
\begin{center}
{\centerline{\epsfig{file=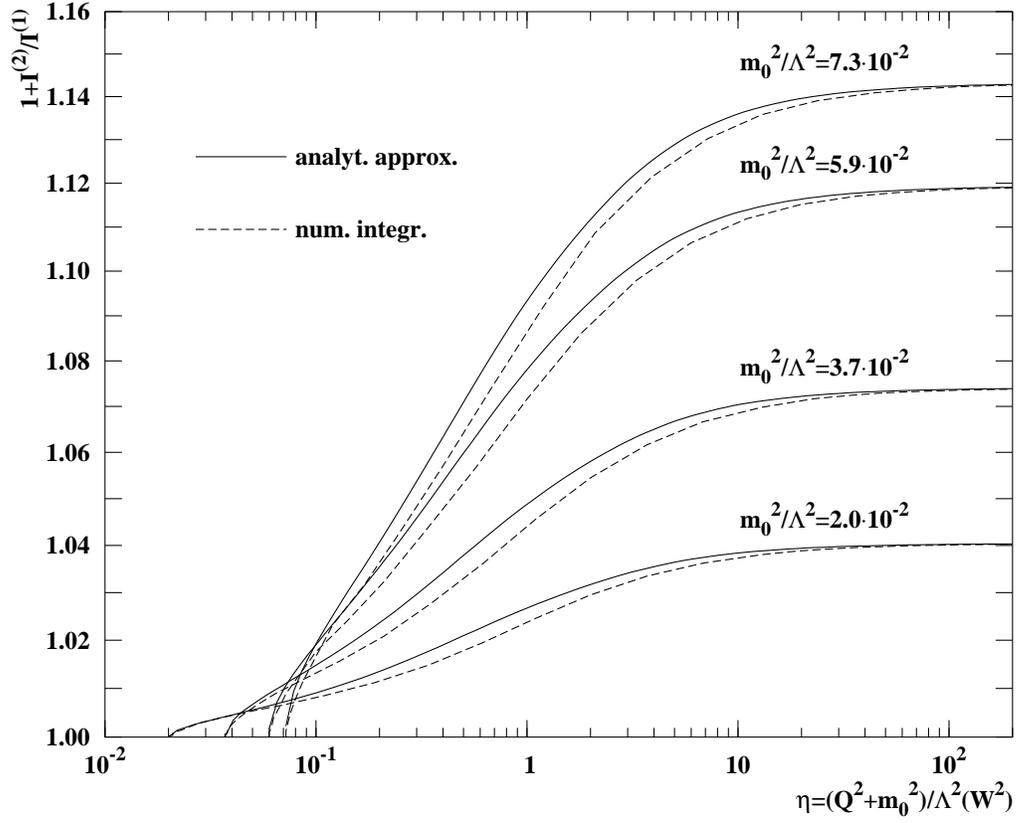,width=14.0cm}}}
\caption{Comparision of the result for the correction factor
$(1+I^{(2)}/I^{(1)})$ obtained by numerical integration with the result from
the analytic approximation  (\protect\ref{2.25}).\label{fig2}}
\end{center}
\end{figure}

\noindent
A comparison of the results of the numerical integration and the
results of the analytic approximation (\ref{2.25}) is shown in
fig.\ref{fig2}. In   the range of  the parameters $\eta$ and
$m_0^2/\Lambda^2(W^2)$ relevant in connection with the experimental
data (compare Section 3), the error induced by employing the
approximation formula (\ref{2.25}) in (\ref{2.22}) and (\ref{2.23})
is less than 0.3 \%. Accordingly, the expression
(\ref{2.23}) for $\sigma_{\gamma^*p} (W^2, Q^2)$ together with (\ref{2.22})
and the
analytical results (\ref{2.24}) and (\ref{2.25}) will form the basis\footnote{
A FORTRAN code for evaluation of $\sigma_{\gamma^*p}$ as a function of
$(W^2,Q^2)$
%can be provided on request.
will be available from  http://www.desy.de/\~{}surrow/gvd.html
} for the
analysis of
the experimental data.

We briefly discuss the function $I (\eta , m^2_0/\Lambda^2 (W^2))$,
in (\ref{2.22}) and (\ref{2.23}) for
various limits of the parameter space that will be relevant for the data
analysis.
First of all, for small values of $m^2_0/\Lambda^2(W^2)$, one
may expand the expression for $I^{(1)} (\eta, m^2_0/\Lambda^2(W^2))$ in
(\ref{2.24}) to yield

\begin{equation}
I^{(1)}
\left( \eta,
                 \frac{
                         m_0^2
                      }{
                         \Lambda^2(W^2)
                      }
\right)=
   {\cal I}_0(\eta)
   +
   {\cal I}_1(\eta) {{m^2_0} \over {\Lambda^2(W^2)}}
   +
   { \rm O }\left( {{m^4_0} \over {\Lambda^4(W^2)}}\right),
                                                    \label{2.26}
\end{equation}
where
\begin{eqnarray}\label{2.27}
\nonumber
{\cal I}_0(\eta)&=&
\frac{1}{
            2\sqrt{1+4\eta}
        }
\ln\frac{
          \eta(1+\sqrt{1+4\eta})
        }{
         -1-3\eta+(1+\eta)\sqrt{1+4\eta}
        },\\
%\nonumber
{\cal I}_1(\eta)&=&
\frac{1}{
          1+4\eta
        }
\left(
      \frac{
             -3
           }{
             1+\eta
           }
      +2{\cal I}_0(\eta)
\right).
\end{eqnarray}

\noindent
A numerical evaluation shows that the term linear in $m^2_0/\Lambda^2 (W^2)$
becomes negligible as long as $m^2_0/\Lambda^2(W^2) \lsim 1$. For
$m^2_0/\Lambda^2(W^2) < 1$, also the correction term
(\ref{2.25}) does not
deviate much from unity, and accordingly, $I (\eta, m^2_0/\Lambda^2(W^2))$
in  (\ref{2.23}), in good approximation, only depends on $\eta$,

\be
I (\eta, m^2_0/\Lambda^2(W^2)) \simeq {\cal I}_0 (\eta).
\label{2.28}
\ee

For a sufficiently smooth $W$ dependence of
$\sigma^{(\infty)} (W^2)$ in (\ref{2.23}), we have approximate scaling of the
virtual-photon proton cross section,

\be
\sigma_{\gamma^*p} (W^2, Q^2) \cong {{\alpha R_{e^+e^-}} \over {3 \pi}}
\sigma^{(\infty)} (W^2) {\cal I}_0(\eta),
\label{2.29}
\ee

\noindent
i.e. in good approximation, the $\gamma^*p$ total cross section only
depends
on the scaling variable $\eta = (Q^2 + m_0)/\Lambda^2 (W^2)$.

It is instructive to consider the limiting cases of small $\eta$ and
large $\eta$ in ${\cal I}_0 (\eta)$. From (\ref{2.27}), one finds

\begin{equation}
{\cal I}_0(\eta) =
\left\{
        \begin{array}{ll}
             \ln(1/\eta) + {\rm O}(\eta\ln \eta),\quad
             &
             \mbox{for } \eta\to\eta_{min}=m_0^2/\Lambda^2(W^2),
           \\[2mm]
             1/(2\eta)+{\rm O}(1/\eta^2),\quad
             &
             \mbox{for } \eta\to\infty.
        \end{array}
\right.
\label{2.30}
\end{equation}

\noindent
The behaviour of $\sigma_{\gamma^*p}(\eta)$ thus changes dramatically, from
a logarithmic one for small $\eta$ to a powerlike one for large $\eta$.
Note that the small-$\eta$ limit besides photoproduction $(Q^2 = 0)$
also includes the limit of fixed $Q^2$, but $\Lambda^2 (W^2)$ sufficiently
large. As $\Lambda^2 (W^2)$ will turn out to increase as a power of $W^2$,
one will be led to the conclusion that at any value of $Q^2$ the
virtual-photon-proton cross section will at sufficiently high energy,
that is for small $\eta$,
exhibit the same smooth
energy dependence that is observed in photoproduction.

\subsection{The photoproduction limit and the significance of
$\sigma^{(\infty)} (W^2)$.}

Evaluating the unpolarized cross section $\sigma_{\gamma^*p}(W^2, Q^2)$
in (\ref{2.23}) for $Q^2 = 0$, or, equivalently, the transverse cross section
(\ref{2.11}), we obtain our result for the cross section of photoproduction,

\be
\sigma_{\gamma p}(W^2) = {{\alpha R_{e^+e^-}} \over {3 \pi}} \sigma^{(\infty)}
(W^2) I \left( \eta (\Lambda^2(W^2),Q^2 = 0), {{m^2_0} \over {\Lambda^2(W^2)}}
\right),
\label{2.31}
\ee

\noindent
where, according to (\ref{2.10}),

\be
\eta (\Lambda^2 (W^2), Q^2 = 0) = {{m^2_0} \over {\Lambda^2(W^2)}},
\label{2.32}
\ee
and $R_{e^+e^-}=2$ is to be inserted.
To proceed, it is suggestive to require duality between the generic
two-gluon-exchange structure of the GVD/CDP contained in (\ref{2.31}) and the
Regge behaviour experimentally verified for photoproduction.
This duality assumption is meant
with respect to Pomeron exchange that dominates photoproduction in the
high-energy limit. Accordingly, we require

\be
\sigma_{\gamma p}(W^2)^{Regge} = {{\alpha R_{e^+e^-}} \over {3 \pi}}
\sigma^{(\infty)} (W^2) I \left( {{m^2_0}\over {\Lambda^2(W^2)}}, {{m^2_0}
\over {\Lambda^2(W^2)}}\right),
\label{2.33}
\ee

\noindent
where the notation $\sigma_{\gamma p} (W^2)^{Regge}$ explicitly displays
the duality hypothesis mentioned above. Solving (\ref{2.33}) for
$\sigma^{(\infty)}
(W^2)$,

\be
\sigma^{(\infty)} (W^2) = {{\sigma_{\gamma p} (W^2)^{Regge}} \over
{{{\alpha R_{e^+e^-}} \over {3 \pi}} I \left( {{m^2_0} \over {\Lambda^2(W^2)}},
{{m^2_0} \over {\Lambda^2(W^2)}}\right)}},
\label{2.34}
\ee

\noindent
allows us to express the virtual-photon-proton cross section (\ref{2.11})
explicitly in terms of\break $\sigma_{\gamma p} (W^2)^{Regge}$,

\be
\sigma_{\gamma^*_{T,L}p} (W^2,Q^2) = \sigma_{\gamma p}(W^2)^{Regge}
{{I_{T,L} \left( \eta ( \Lambda^2 (W^2), Q^2), {{m^2_0} \over {\Lambda^2(W^2)}}
\right)} \over {I \left( {{m^2_0} \over {\Lambda^2 (W^2)}}, {{m^2_0}
\over {\Lambda^2 (W^2)}} \right)}}.
\label{2.35}
\ee

Concerning the relation (\ref{2.34}), it seems appropriate to remind ourselves
of the meaning of $\sigma^{(\infty)} (W^2)$. According to (\ref{2.9}),
$\sigma^{(\infty)} (W^2)$ denotes the limiting behaviour of the colour-dipole
cross section, $\sigma_{(q \bar q)p} (\vec r^{~2}_\perp, z, W^2)$, both for
$r_\perp \to \infty$ with $\Lambda^2(W^2)$ fixed, and for $\Lambda^2 (W^2)
\to \infty$ with $r_\perp$ fixed. The $W$ dependence of $\sigma^{(\infty)}
(W^2)$, as a consequence of the (logarithmic) increase with energy of the
denominator in (\ref{2.34}), in general will deviate from the one of
$\sigma_{\gamma p} (W^2)^{Regge}$. This is not unexpected; the energy
dependence of the colour-dipole cross section a priori need not
coincide with the energy dependence characteristic for ordinary hadron-hadron
interactions.

In connection with the  energy dependence and the conceptual meaning of
$\sigma^{(\infty)} (W^2)$, a brief discussion of the $\rho^0, \omega,
\phi$-dominance \cite{Sak} approximation for the virtual-photon-proton and,
in particular, the
photoproduction cross section will be helpful. Returning to (\ref{2.11}), and
ignoring the off-diagonal transitions in (2.13), we approximate the
integral (\ref{2.13}) by its integrand to obtain

\be
\sigma_{\gamma^*_T p} (W^2, Q^2) = {{\alpha R_{e^+e^-}} \over {3 \pi}}
{{\Delta M^2_\rho} \over {M^2_\rho}} {{M^4_\rho} \over {(Q^2 + M^2_\rho)^2}}
\sigma^{(\infty)} (W^2).
\label{2.36}
\ee

\noindent
In (\ref{2.36}), we made the simplifying assumption of flavour-independent
equal masses,
$M^2 \equiv M^2_\rho$, and equal level spacings, $\Delta M^2 \equiv
\Delta M^2_\rho$, for the dominant vector mesons, $\rho^0, \omega$ and $\phi$.
Moreover, by assuming $R_{e^+ e^-} = 2$, we ignore more
massive vector-meson flavours, such as $J/\psi$, etc. The connection of (\ref{2.36})
with $\rho^0, \omega,
\phi$-dominance for (virtual-) photon-hadron interactions becomes
explicit by introducing the photon-vector-meson coupling strengths via
quark-hadron duality \cite{dual},

\be
{{\alpha R_{e^+e^-}} \over {3 \pi}} {{\Delta M^2_\rho} \over {M^2_\rho}}
= \sum_{V = \rho^0, \omega, \phi} {{\alpha \pi} \over {\gamma^2_V}},
\label{2.37}
\ee

\noindent
as well as the identification of $\sigma^{(\infty)} (W^2)$ with the
total cross section of vector-meson-proton scattering,

\be
\sigma^{(\infty)} (W^2) = \sigma_{Vp} (W^2).
\label{2.38}
\ee

\noindent
As a consequence of the simplification of flavour independence,
$\sigma_{Vp} (W^2)$  denotes a weighted average of the $(\rho^0 p),
(\omega p)$ and $(\phi p)$ cross sections. As the $(\rho^0 p)$ and
$(\omega p)$ cross sections agree with each other, and the $\phi$
contribution is suppressed by the smaller coupling to the photon and by
the smaller $(\phi p)$ cross section \cite{SaSchi},
the weighted average in (\ref{2.38}) may
approximately be identified with the  $(\rho^0 p)$ cross section,

\be
\sigma^{(\infty)} (W^2) \cong \sigma_{\rho p} (W^2).
\label{2.39}
\ee

\noindent
The couplings in (\ref{2.37}), by definition, denote the coupling
strengths of the
photon to the vector mesons, $V = \rho^0, \omega, \phi$, as measured in
$e^+ e^-$ annihilation by the integrals over the corresponding vector-meson
peaks,

\be
{{\alpha \pi} \over {\gamma^2_V}} = {1 \over {4 \pi^2 \alpha}} \sum_F
\int \sigma_{e^+e^- \to V \to F} (s) ds,
\label{2.40}
\ee

\noindent
or equivalently, by the vector-meson widths,

\be
\Gamma_{V \to e^+e^-} = {{\alpha^2 M^2_V} \over {12 (\gamma^2_V/4 \pi)}}.
\label{2.41}
\ee

\noindent
Upon inserting the quark-hadron-duality
relation (\ref{2.37}) and the hadronic
vector-meson-proton cross section (\ref{2.39}) into (\ref{2.36}), we obtain the
$\rho^0, \omega, \phi$-dominance
prediction for the transverse virtual-photon-proton cross section. It exhibits
the well-known violent disagreement with experiment for $Q^2 > 0$, even
though $\rho^0, \omega, \phi$-dominance
yields a reasonable approximation for photoproduction \cite{Stod}.
Dropping the above simplification of flavour independence, it reads
\cite{Stod}
\be
\sigma_{\gamma p} (W^2) = \sum_{V = \rho^0, \omega, \phi} {{\alpha \pi}
\over {\gamma^2_V}} \sigma_{V p} (W^2)={{\alpha \pi}\over {\gamma^2_\rho}}
\sigma_{\rho p}\left(1+{1\over 9} +{2\over 9}\cdot {1 \over 2} \right) ,
\label{2.42}
\ee

\noindent
where the relative weight of the $w$ and $\phi$ contributions is determined
by the quark content of their wave functions, and $\sigma_{\phi p}\cong
(1/2)\sigma_{\rho p}$ is used. Numerically, from (\ref{2.41}), by inserting
$\Gamma_{V \to e^+e^-}\cong 6.5{\rm keV}$, one finds $\gamma^2_\rho /4\pi
\cong 0.53$, and accordingly (\ref{2.42}) yields
\be
\sigma_{\gamma p} (W^2) = (1/240)\sigma_{\rho p}(W^2).
\label{2.42a}
\ee
This relation will be used in Section 3.3. It is of approximate validity. A
careful analysis at energies around $W\cong 3 {\rm GeV}$ revealed that the
right-hand side in (\ref{2.42a}) yields 78 \% of $\sigma_{\gamma p}$
\cite{SaSchi}.

Even though the above exposition of how the $\rho^0, \omega, \phi$-dominance
approximation is contained in the GVD/CDP may be useful in its own right,
it has been our main concern in this section  to illuminate the meaning of
$\sigma^{(\infty)} (W^2)$. In general, $\sigma^{(\infty)} (W^2)$ in
(\ref{2.34}) differs conceptually, and in its energy dependence, from a
vector-meson-proton cross section. It is in  the $\rho^0, \omega, \phi$-dominance
approximation (\ref{2.36})  that an identification of
$\sigma^{(\infty)}(W^2)$ with the
hadronic cross section, $\sigma_{\rho p} (W^2)$ in (\ref{2.39}),
becomes justified.
A strict validity of (\ref{2.39}), however, when inserted into (\ref{2.31}),
requires
$\Lambda^2 (W^2)$ to be an energy-independent constant. Such a requirement,
in turn, implies the energy dependence of $\sigma_{\gamma^*p}(W^2,Q^2)$
to be identical for all $Q^2$,
in gross disagreement with the experimental results from HERA
\cite{data1,data2}.

\subsection{A reference to pre-QCD off-diagonal generalized vector dominance}

As strongly emphasized before \cite{Cvetic}, and explicitly displayed in
(\ref{2.5}), in the GVD/CDP, it is the generic two-gluon-exchange
structure of the $(q \bar q) p$ interaction that leads to the characteristic
difference in sign between, and the necessary cancellation of diagonal
and off-diagonal contributions to the virtual-forward-Compton-scattering
amplitude. The difference in sign corresponds to destructive interference
between hadron-production amplitudes induced by different masses of the
$q \bar q$ states the incoming photon dissociates into. The destructive
interference is a necessity \cite{FRS} for the convergence of the mass dispersion
relations (\ref{2.13}) to (\ref{2.16}), or, in other words, the
consistency of scaling in $e^+ e^-$ annihilation into hadrons with the
GVD picture of DIS at low x. Off-diagonal transitions in the mass
dispersion relation, in order to simplify the formalism, were frequently
ignored \cite{SaSchi,SpSchi}
in the past, at the expense of introducing an ad hoc effective
$1/M^2$ decrease of the $(q \bar q) p$ strong-interaction cross section.
This approximation is confronted with consistency problems \cite{FRS}, and an
approach that does not rely on the diagonal approximation is
preferable right from the outset.

The necessary cancellation in the virtual-forward-Compton amplitude
between diagonal and off-diagonal transitions was anticipated \cite{FRS}
during the pre-QCD era. We indicate how an approximate evaluation of
the GVD/CDP indeed coincides\footnote{
For the connection between the GVD/CDP and off-diagonal GVD, compare also
\cite{Forshaw} and \cite{Frankfurt}
} with the pre-QCD formulation of off-diagonal GVD.

We concentrate on the transverse part of the virtual-photon-proton
cross section in (\ref{2.11}) and consider the off-diagonal term in the
mass dispersion relation (\ref{2.13}). In order to find an approximate
evaluation of the off-diagonal contribution to the integral in (\ref{2.13}),
we start by tentatively putting $M^{\prime 2} = M^2$ in the denominator
of (\ref{2.13}). Under this assumption, the integration over $dM^{\prime 2}$
can be easily carried out by employing the integral relations for $\omega
(M^2, M^{\prime 2}, \Lambda^2 (W^2))$ in (\ref{2.17}) and (\ref{2.18}).
One notes that the result of this integration is identically reproduced
by replacing $M^{\prime 2}$ by $M^{\prime 2} = M^2 + \Lambda^2 (W^2)$
in the multiplicative factor in front of $\omega (M^2, M^{\prime 2},
\Lambda^2 (W^2))$ in (\ref{2.13}) prior to integrating over $dM^{\prime 2}$.
Returning to the correct off-diagonal term in (\ref{2.13}) by dropping the
simplifying assumption of $M^{\prime 2} = M^2$ in the denominator in
(\ref{2.13}), we now replace $M^{\prime 2}$ in the factor multiplying
$\omega (M^2, M^{\prime 2}, \Lambda^2 (W^2))$ by the more general mean
value

\be
M^{\prime 2} = M^2 + {{\Lambda^2 (W^2)} \over {1 + 2 \delta_T}},
\label{2.43}
\ee

\noindent
that contains the parameter $\delta_T$. The parameter $\delta_T$ is to
be chosen such that the off-diagonal integral is properly reproduced
in the sense of a mean-value evaluation of this integral. With the
substitution (\ref{2.43}), as specified, and upon integration over
$dM^{\prime 2}$, using (\ref{2.17}), the integral (\ref{2.13}) becomes

\be
I_T \left(\eta, {{m^2_0} \over {\Lambda^2(W^2)}}\right) \simeq
\int^\infty_{m^2_0} dM^2 \left[ {{M^2} \over {(Q^2 + M^2)^2}} -
{{M^2 - \delta_T {{\Lambda^2 (W^2)} \over {1 + 2 \delta_T}}} \over {(Q^2+M^2)
(Q^2 + M^2 + {{\Lambda^2 (W^2)} \over {1 + 2 \delta_T}}}}\right].
\label{2.44}
\ee

\noindent
With (\ref{2.44}), and using the quark-hadron-duality relation (\ref{2.37}), as
well as the identification (\ref{2.39}), the cross section resulting from
(\ref{2.11}),

\begin{eqnarray}\label{2.45}
\sigma_{\gamma^*_Tp} (W^2, Q^2) & &\hspace*{-0.5cm} =
\sum_{V = \rho^0, \omega, \phi} {{\alpha \pi} \over {\gamma^2_V}}
\sigma_{V p} (W^2) {{m^2_\rho} \over {\Delta m^2_\rho}}
\nonumber \\
& &\hspace*{-0.5cm} \times \int^\infty_{m^2_0} dM^2
\left[ {{M^2} \over {(Q^2 + M^2)^2}} - {{M^2 - \delta_T {{\Lambda^2 (W^2)}
\over {1 + 2 \delta_T}}} \over {(Q^2 + M^2) (Q^2 + M^2 + {{\Lambda^2(W^2)}
\over {1 + 2 \delta_T}}}} \right],
\end{eqnarray}

\noindent
coincides\footnote{In ref. \cite{FRS}, compare (4) upon substituting (2) and (6),
and replace the sum in (4) by an integral.} with the one in ref. \cite{FRS} in
the approximation that $\Lambda^2$ and $\delta_T$ are treated as
appropriately chosen constants.

The original derivation \cite{FRS} that led to (\ref{2.45}) was based on an
infinite series of discrete vector-meson states. The opposite signs of
diagonal and off-diagonal transitions were located at the $\gamma^*
(q \bar q)$-transition vertices, as, e.g., had been suggested by bound-state
quark-model calculations \cite{BJK}. It is amusing to note that the
anticipated structure, in the framework of QCD, now finds an entirely
different justification: the origin of the sign difference has shifted
from the $\gamma^* (q \bar q)$ vertices to the generic structure of the
two-gluon-exchange amplitude in the purely hadronic $(q \bar q)p$ interaction.

\section{The generalized vector dominance/colour-dipole picture confronting
the experimental data on $\sigma_{\gamma^* p} (W^2, Q^2)$.}
\setcounter{equation}{0}

In confronting the theoretical results from Section 2 with the experimental
data, we will follow the strategy employed in our recent communication
\cite{Surrow}.
In Section 3.1, accordingly, the prediction (\ref{2.29})
of scaling for the $\gamma^* p$ total cross section,
\begin{equation}\label{3.1}
\sigma_{\gamma^* p} (W^2, Q^2) \simeq \sigma_{\gamma^* p} (\eta)
\end{equation}
will be tested in a model-independent approach. In Section 3.2,
the empirical validity of the
functional dependence of $\sigma_{\gamma^* p}(\eta)$ on $\eta$
in the GVD/CDP
given in (\ref{2.22}) to (\ref{2.25})
will be investigated.

\subsection{Low-$x$ scaling in $\gamma^* p$ total cross sections}

The scaling variable $\eta$ was defined by (\ref{2.10}) as the ratio of
$Q^2 + m^2_0$ over $\Lambda^2 (W^2)$. According to (\ref{2.8}),
$\Lambda(W^2)$ determines the magnitude of the (two-dimensional)
momentum transfer $|\vec l_\bot |$ to the quark and antiquark. As a
consequence, $\Lambda (W^2)$ also determines the magnitude of the
final-state $q \bar q$ masses, $M^\prime$, that can be reached from
a given $q \bar q$ mass, $M$, in the initial state. This
interpretation suggests that $\Lambda^2 (W^2)$ be an increasing function of
the energy, $W$. We will adopt a power-law ansatz,
\begin{equation}\label{3.2}
\Lambda^2 (W^2) = C_1 (W^2 + W^2_0)^{C_2},
\end{equation}
and, alternatively, a logarithmic one,
\begin{equation}\label{3.3}
\Lambda^2 (W^2) = C^\prime_1 \ln \left( \frac{W^2}{W^{\prime 2}_0} +
C^\prime_2 \right) .
\end{equation}
Altogether, the scaling variable $\eta$ depends on $m^2_0$ and the constants
$C_1 , W^2_0 , C_2$ (or, alternatively, $C^\prime_1 , W^{\prime 2}_0 , C^\prime
_2$) to be determined by a fit to the data based on the scaling conjecture
(3.1).
In the model-independent test of scaling, no specific functional
dependence of the total cross section (3.1)
on $\eta$ is assumed. Accordingly, in the model-independent analysis,
the parameter $C_1$ (or, alternatively, the parameter $C^\prime_1$)
remains undetermined. A change of $C_1$ (or $C^\prime_1$) amounts to a
rescaling of $\eta$ to $C^{-1}_1 \eta$ (or $C^{\prime -1}_1 \eta$), and,
accordingly, the absolute value of $C_1$ (or $C^\prime_1$) is irrelevant
for the existence of a scaling behaviour for the $\gamma^* p$ total
cross section.
For the analysis and the representation of the data, we will use a value of $C_1$ (or
$C^\prime_1$) that coincides with, or is in the vicinity of the value
to be determined in the fit to the data based on the GVD/CDP in
Section 3.2.

Technically, the empirical test of the scaling law (3.1)
is carried out as follows. Without loss of generality, we assume that the
conjectured scaling curve for $\sigma_{\gamma^* p} (\eta)$ may be
represented by a piecewise linear function of $\eta$. This assumption
allows us to perform a fit to the data that determines the parameters
$m^2_0, W^2_0 , C_2$ simultaneously with the values of the
piecewise linear function $\sigma_{\gamma^* p} (\eta)$ at a number of points
$\eta_i (i = 1, ..., N)$  of the variable $\eta$.

\begin{figure}[ht]
\vspace*{-1.5cm}
\begin{center}
{\centerline{\epsfig{file=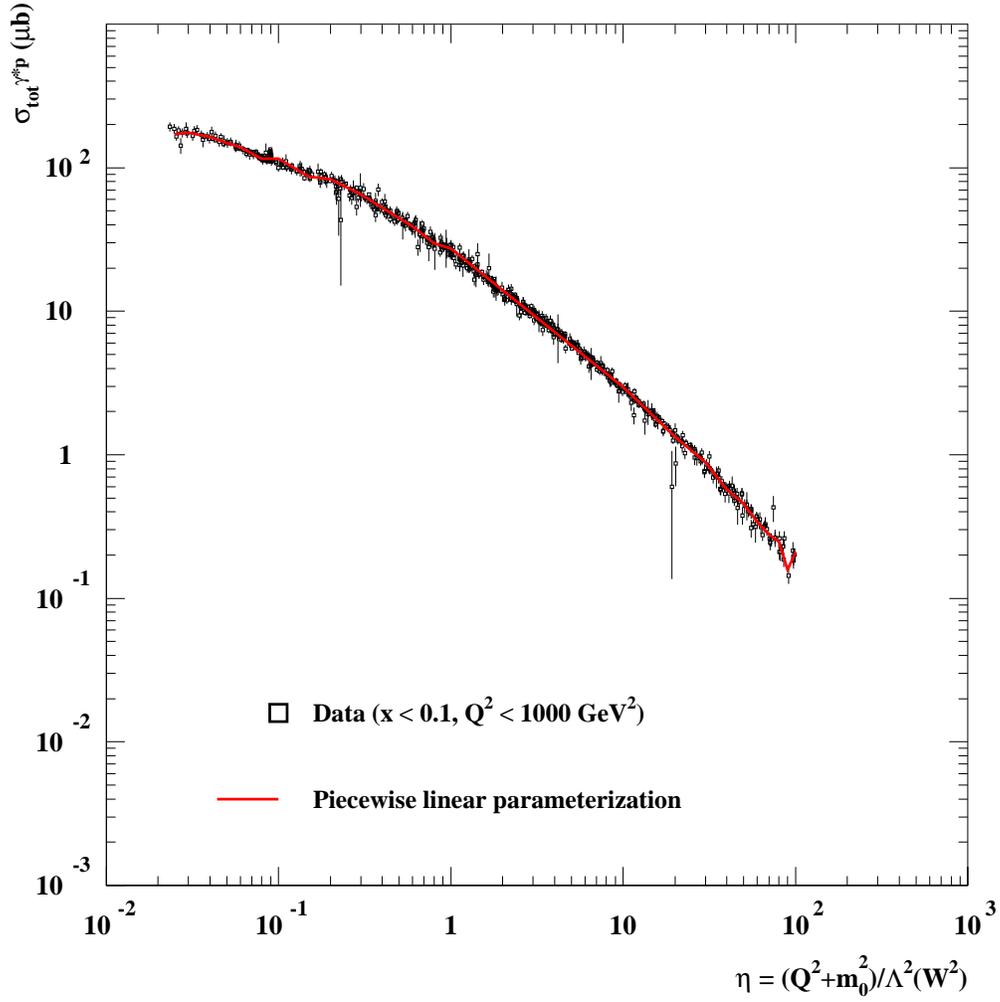,width=14.0cm}}}
\caption{The experimental data for $\sigma_{\gamma^* p} (W^2,Q^2)$ for
$x \simeq Q^2/W^2 < 0.1$ vs. the low-x scaling variable
$\eta = (Q^2 + m^2_0) / \Lambda^2 (W^2)$.\label{fig3}}
\end{center}
\end{figure}

In fig.\ref{fig3}, we show the result of the model-independent analysis. For $C_1
(C^\prime_1)$ the numerical value of $C_1 = 0.34$ ($C^\prime_1 = 1.64\ {\rm
GeV}^2$)
was chosen. As seen in fig.\ref{fig3}, upon imposing the kinematic restriction of
$x \le 0.1$ and $Q^2 \le 1000{\rm GeV}^2$, all available experimental data
\cite{data1,data2,data3,data4,data5}
on photo- and electroproduction are indeed seen to lie on a smooth curve that,
for technical reasons, is approximated by the piecewise linear fit function.
The parameters obtained from the fit, using the power-law ansatz (3.2),
are given by
\begin{eqnarray}\label{3.4}
m^2_0 & = & 0.125 \pm 0.027 {\rm GeV}^2 ,\nonumber  \\
C_2 & = & 0.28 \pm 0.06 , \\
W^2_0 & = & 439 \pm 94 {\rm GeV}^2 , \nonumber
\end{eqnarray}
with $\chi^2$ per degree of freedom, $\chi^2$/ndf=1.15.
For the logarithmic ansatz, we obtained
\begin{eqnarray}\label{3.5}
m^2_0 & = & 0.12\pm 0.04{\rm GeV}^2, \nonumber \\
C^\prime_2 & = & 3.5\pm 0.6,\\
W^{\prime 2}_0 & = &  1535 \pm 582{\rm GeV}^2,\nonumber
\end{eqnarray}
with $\chi^2$/ndf=1.18.

We note that an analogous procedure, applied to the experimental data without
a restriction on $x$, does {\it not} lead to a universal curve. Likewise,
restricting oneself to only those data points that belong to
$x > 0.1$, {\it no} universal curve is obtained either; the fitting
procedure leads to entirely unacceptable results on the quality of the fit
as quantified by the value of $\chi^2$ per degree of freedom.

The model-independent phenomenological analysis thus reveals that the scaling
behaviour of the virtual-photon-proton cross section derived from the
GVD/CDP is indeed borne out by the experimental data. This result by itself
does not allow one to conclude that also the specific functional dependence
on $\eta$
of the GVD/CDP, or on $W^2$ and $Q^2$ as given in Section 2.2,
holds for the data on the $\gamma^* p$ total cross section. To
investigate this question, we turn to Section 3.2.

\subsection{Testing the $\eta$ dependence of $\sigma_{\gamma^* p}$
in the GVD/CDP}

Upon replacing $\sigma^{(\infty)}(W^2)$ in (\ref{2.23})
by the Regge-parameterization of the total photoproduction cross section
according to (\ref{2.34}), we have
\begin{equation}\label{3.6}
\sigma_{\gamma^* p} (W^2 , Q^2) = \sigma_{\gamma p} (W^2)^{Regge}
\frac{I(\eta (\Lambda^2 (W^2), Q^2), \frac{m^2_0}{\Lambda^2(W^2)})}
{I ( \frac{m^2_0}{\Lambda^2(W^2)}, \frac{m^2_0}{\Lambda^2(W^2)})}.
\end{equation}
The analytical results for $I(\eta , \frac{m^2_0}{\Lambda^2(W^2)})$ to be
employed in the fits are given in (\ref{2.22})
with (\ref{2.24}) and (\ref{2.25}). For $m_0^2/\Lambda^2(W^2)\ll 1$, we have
(approximate) scaling, compare (\ref{2.28}) and (\ref{2.30}).

In (\ref{3.6}), for the photoproduction cross section, we use the
parameterization
\begin{equation}\label{3.7}
\sigma_{\gamma p}(W^2)^{Regge}=A_R(W^2)^{\alpha_R-1}+A_P(W^2)^{\alpha_P-1},
\end{equation}
where $W^2$ is to be inserted in units of $\rm GeV^2$ and \cite{surrowZEUS}
\begin{eqnarray}
A_R & = & 145.0 \pm 2.0\; {\rm \mu b} , \nonumber \\
\alpha_R & = & 0.5,    \label{3.8} \\
A_P &  = & 63.5 \pm 0.9\; {\rm \mu b} , \nonumber \\
\alpha_P & = & 1.097 \pm 0.002. \nonumber
\end{eqnarray}
For a test of the empirical validity of the GVD/CDP formula (\ref{3.6}), one
may evaluate (\ref{3.6}) for the power-law ansatz for $\Lambda^2(W^2)$ in
(\ref{3.2}), or the logarithmic one in (\ref{3.3}), using the parameters
(\ref{3.4}) and (\ref{3.5}) from the model-independent fit, and determine
$C_1(C_1^\prime)$ by a fit of (\ref{3.6}) to the experimental data
for $\sigma_{\gamma^* p}(W^2,Q^2)$.

\begin{figure}[ht]
\vspace*{-0.5cm}
\begin{center}
{\centerline{\epsfig{file=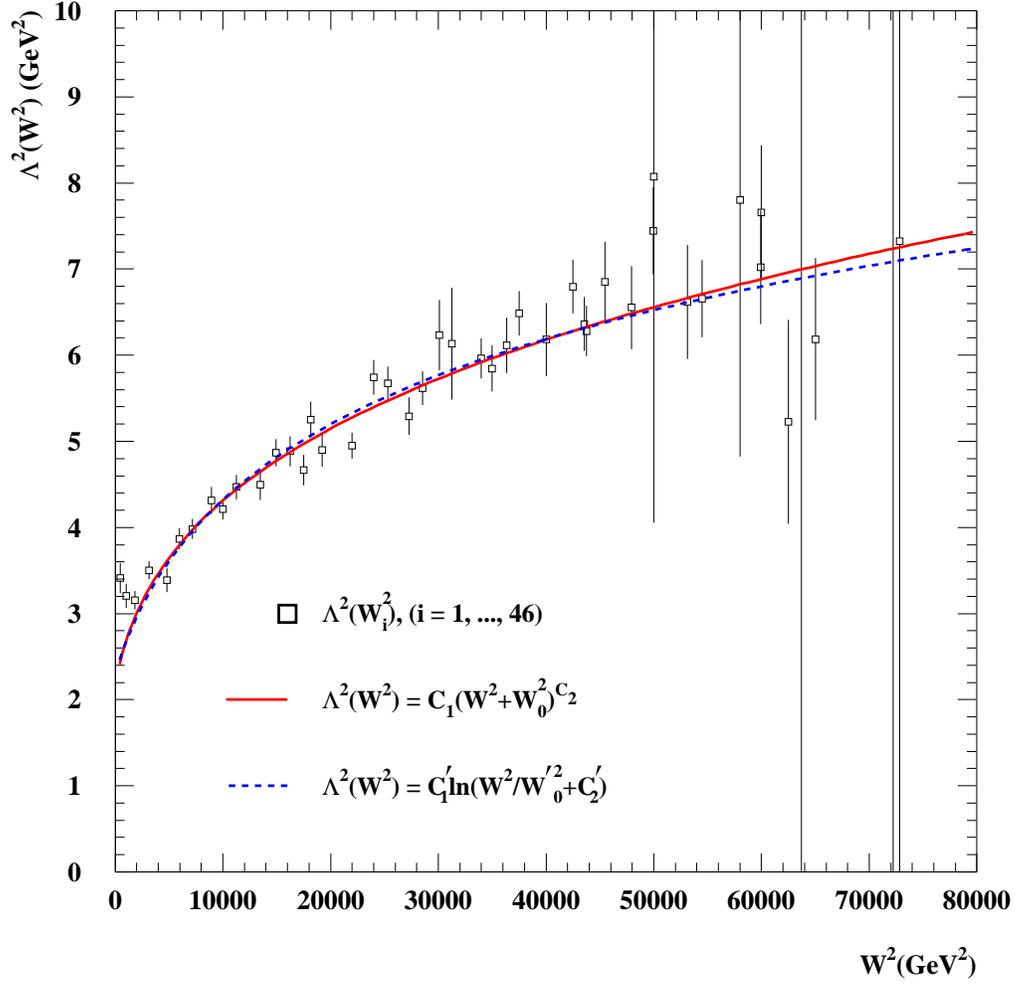,width=14.0cm}}}
\caption{The dependence of $\Lambda^2$ on $W^2$, as determined by
a fit of the GVD/CDP
predictions for $\sigma_{\gamma^* p}$ to the experimental data.\label{fig4}}
\end{center}
\end{figure}

The  alternative approach, actually employed in our analysis, is as follows.
Rather than  relying on the functional  form of, and the values of  the
parameters in $\Lambda^2(W^2)$ from the model-independent analysis, we only
assume   that $\Lambda^2(W^2)$ can be described  by a smooth piecewise
linear function of $W^2$. The  fit of (\ref{3.6}) to the  data for
$\sigma_{\gamma^* p}(W^2,Q^2)$ then is to determine $m_0^2$, as well as the
values of $\Lambda^2(W^2)$ at a set of chosen values, $W^2_i$, for
$i=1,{\ldots}, N $. The values of $\Lambda^2(W^2_i)$ obtained in our fit
(for $i=1,{\ldots}, 46 $) under the restriction of $x\leq 0.01$ and $Q^2 \leq
100 GeV^2$ are shown in fig.\ref{fig4}. This fitting procedure,  with an acceptable
$\chi^2/ndf=1.15$, provides the most direct empirical verification of the
$Q^2$ dependence of the GVD/CDP. At any energy,  $W_i$, the $Q^2$
dependence, by our fit,
is indeed verified  to be described by (\ref{2.13}) to (\ref{2.16}), or
rather (\ref{2.24}) and (\ref{2.25}), upon inserting the appropriate value
of $\Lambda^2(W^2_i)$ from fig.\ref{fig4}.

It is in a second step  that we now  assume the powerlike and the
logarithmic  analytical  form,   respectively, for $\Lambda^2(W^2)$ in
(\ref{3.2}) and (\ref{3.3}), in order to fit (\ref{3.6}) to the experimental
data for the $\gamma^*p$ interaction again. The resulting curves for
$\Lambda^2(W^2)$ are also displayed in fig.\ref{fig4}, and are seen to provide a
good representation of the results for $\Lambda^2(W^2_i)$.
The fit parameters, in distinction to (\ref{3.4}) and (\ref{3.5}),
now include the absolute normalization $C_1$ $(C_1^\prime)$ of the scaling
variable $\eta$. The fitted parameters are given by
\begin{eqnarray} \label{3.9}
m_{0}^{2} & = & 0.16 \pm 0.01\; {\rm GeV^{2}}, \nonumber \\
C_{1}     & = & 0.34 \pm 0.05,             \\
C_{2}     & = & 0.27 \pm 0.01 , \nonumber \\
W_{0}^{2} & = & 882 \pm 246\; {\rm GeV^{2}},  \nonumber
\end{eqnarray}
with $\chi^{2}/ndf=1.15$ for the power-law ansatz, and by
\begin{eqnarray}\label{3.10}
m_{0}^{2} & = & 0.157 \pm 0.009\; {\rm GeV^{2}}, \nonumber \\
C_1^\prime     & = & 1.64 \pm 0.14{\rm GeV^{2}}, \\
C_2^\prime     & = & 4.1 \pm 0.4 , \nonumber \\
{W^\prime}^2_0 & = & 1015 \pm 334\; {\rm GeV^{2}},  \nonumber
\end{eqnarray}
with $\chi^{2}/ndf=1.19$ for the logarithmic one. Since both, the model-independent
fit and the one based on the GVD/CDP, describe the experimental data, the
parameters in $\eta\Big(\Lambda^2(W^2),Q^2\Big)$ resulting from the
different fit procedures must be consistent with each other. This is the
case, compare (\ref{3.9}) and (\ref{3.10}) with (\ref{3.4}) and (\ref{3.5}),
respectively.

It is worth stressing at this point that $\Lambda^2(W^2)$, shown in
fig.\ref{fig4}, not only yields the denominator of the scaling variable $\eta$.
According to (\ref{2.9}), $\Lambda^2(W^2)$ directly determines the
energy dependence of the colour-dipole cross section in the limit of
$\Lambda^2(W^2)\vec r^{~2}_\perp\to 0$, that is the limit of sufficiently
small interquark separation in the colour dipole and for non-asymptotic
energies.

\begin{figure}[ht]
%\vspace*{-1.5cm}
\begin{center}
{\centerline{\epsfig{file=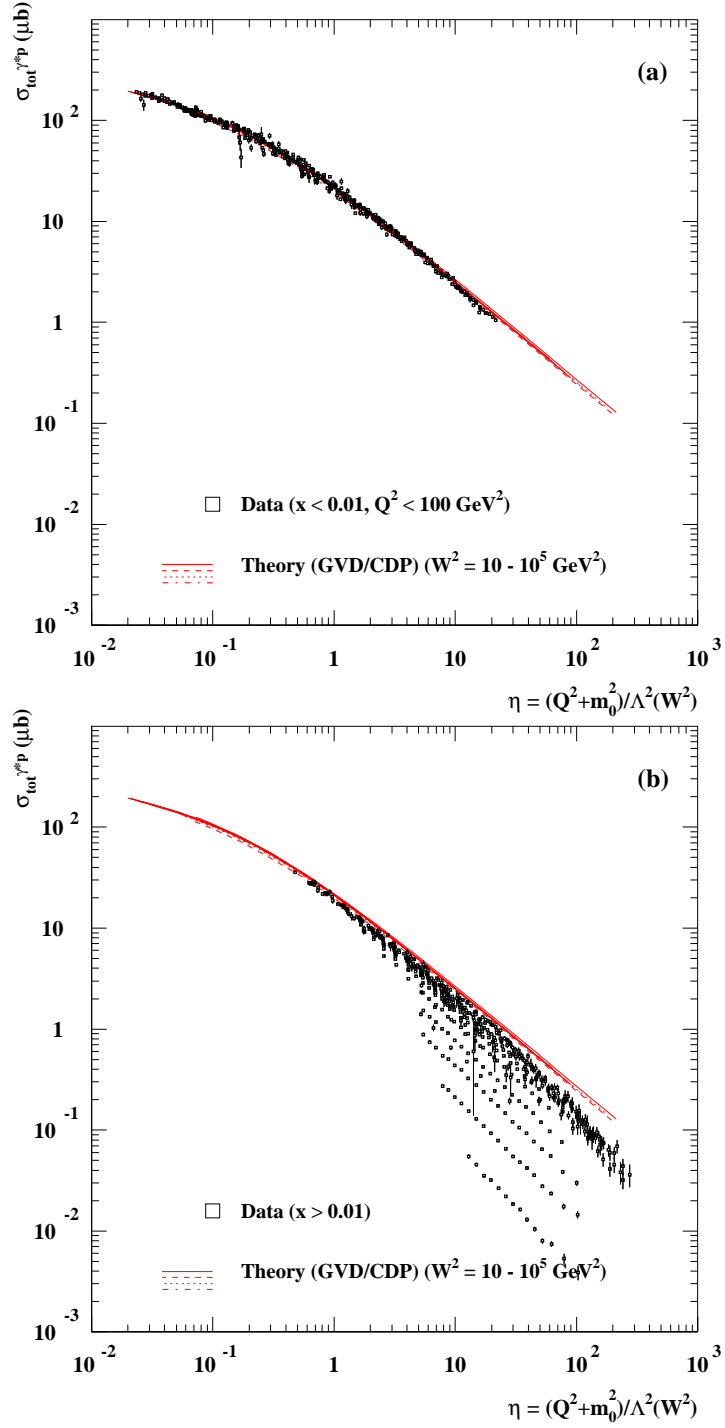,width=10.0cm}}}
\caption{The GVD/CDP scaling curve for $\sigma_{\gamma^* p}$ compared with the
experimental data a) for $x < 0.01$, b) for $x > 0.01$.\label{fig5}}
\end{center}
\end{figure}

In fig.\ref{fig5}, we show the explicit comparison of the experimental data
for $\sigma_{\gamma^*p}(W^2,Q^2)$ as a function of $\eta$ with the
theoretical results of the GVD/CDP. The (approximate) coincidence of the
theoretical predictions over a wide  range of $W^2$, from $W^2\approx 10
\mbox{GeV}^2$ to $W^2\approx 10^5\mbox{GeV}^2$, demonstrates the scaling
property of the theory. As shown in fig.\ref{fig5}a, with the restrictions
$x<0.01$ and $Q^2<100 GeV^2$ imposed on the data (as in the above fit),
there is good agreement between theory and experiment. In fig.\ref{fig5}b, we
show the deviations between  theory and experiment, occurring when data for
$x\geq 0.01$ are taken into account exclusively.

\begin{figure}[ht]
\vspace*{-1.5cm}
\begin{center}
{\centerline{\epsfig{file=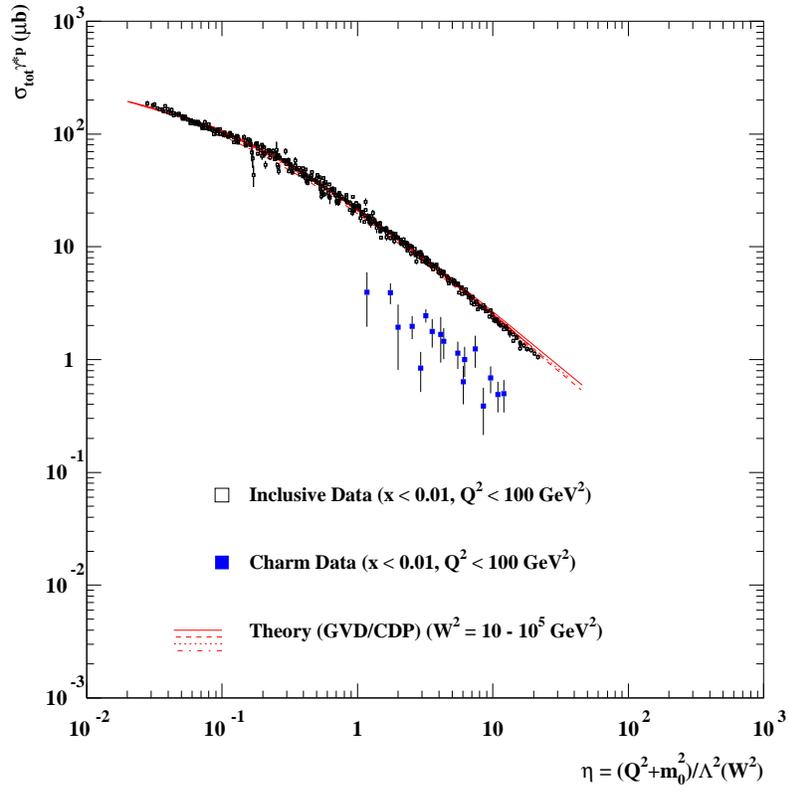,width=11.0cm}}}
\caption{The cross section for charm production, $\sigma_{\gamma^*
p}^{charm}$, in addition to the total cross section, $\sigma_{\gamma^* p}$, as
a function of $\eta$.\label{fig6}}
\end{center}
\end{figure}

One may wonder about the influence of the charm contribution on the total
cross section  with respect to the scaling behaviour in $\eta$. A priori, one
may expect charm production, when analysed by itself, to lead to a different
mass scale, $m^2_{charm}>m^2_0$, in $\eta$. In fig.\ref{fig6}, we have
plotted the experimental data \cite{charm} for
$\sigma_{\gamma^*p}^{charm}(W^2,Q^2)$ against $\eta$, in addition to
the total cross
section, $\sigma_{\gamma^*p}(W^2,Q^2)$. The
charm-production data contribute roughly 30 \% to $\sigma_{\gamma^*p}(\eta)$,
but otherwise show approximately the same dependence on $\eta$ as
observed for $\sigma_{\gamma^*p}(\eta)$. Note that for the charm data, $Q^2\gsim
10\mbox{GeV}^2$, such that it is fairly irrelevant whether
$\eta=(Q^2+m_0^2)/\Lambda^2(W^2)$ or $\eta=(Q^2+m_{charm}^2)/\Lambda^2(W^2)$
is used as a scaling variable. Clearly, if precise data on charm production will
be analysed with respect to their scaling properties, one expects to arrive
at $m^2_{charm}$ replacing $m_0^2$ in $\eta$. This is irrelevant for the
total cross section, however, since with smaller values of $Q^2$, charm
production soon becomes a minor contribution to $\sigma_{\gamma^*p}(W^2,Q^2)$.

\begin{figure}[ht]
\vspace*{-1.5cm}
\begin{center}
{\centerline{\epsfig{file=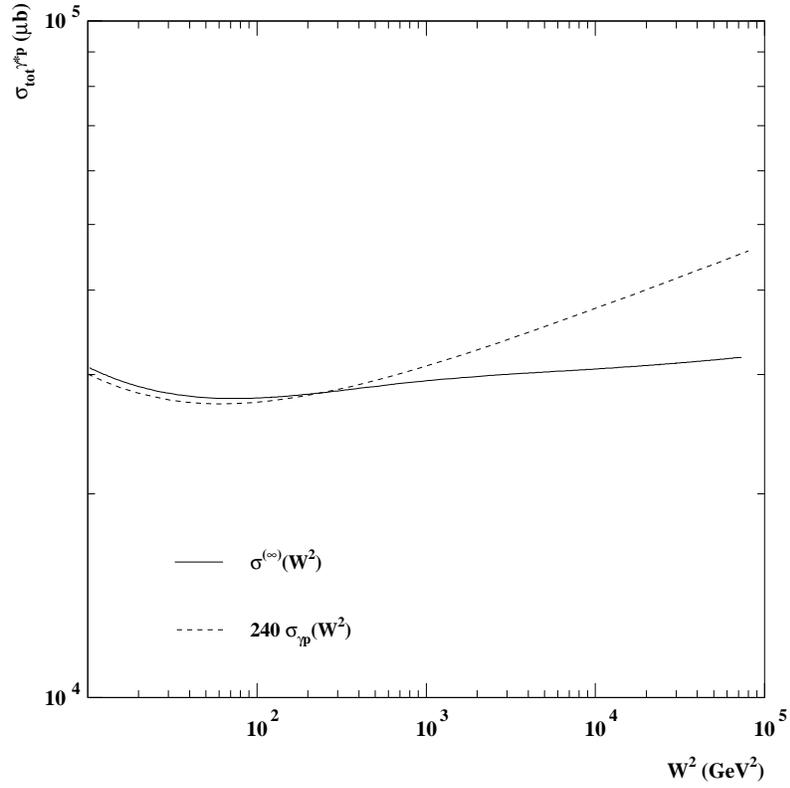,width=11.0cm}}}
\caption{ The asymptotic value $\sigma^{(\infty)}(W^2)$ of the
colour-dipole-proton cross section.
For comparision, also the photoproduction cross section, multiplied by the
factor 240 from $\rho^0,\omega,\phi$ dominance (\protect\ref{2.42a}),
is shown.\label{fig7}}
\end{center}
\end{figure}

As noted, the theoretical prediction (\ref{3.6}) is based on the
replacement of the asymptotic colour-dipole cross section,
$\sigma^{(\infty)}(W^2)$, in (\ref{2.23}) in terms of photoproduction according to the
duality relation (\ref{2.34}). In fig.\ref{fig7}, we represent
$\sigma^{(\infty)}(W^2)$ as a function of $W^2$, calculated according to
(\ref{2.34}) by inserting the Regge fit (\ref{3.7}) for $\sigma_{\gamma
p}(W^2)^{Regge}$ and $\Lambda^2(W^2)$ from (\ref{3.2}) with the parameters
(\ref{3.9}). We also show the cross section of photoproduction scaled by the
factor 240 according to the $\rho^0, \omega,\phi$-dominance prediction
(\ref{2.42a}). At low energies, $\sigma^{(\infty)}(W^2)$ is well
approximated by the scaled photoproduction cross section. The energy
dependence at low energies is dominated by the Regge term in (\ref{3.7})
proportional to $(W^2)^{\alpha_R-1}$. The absolute magnitude of
$\sigma^{(\infty)}(W^2)$ turned out to be somewhat larger than
$\sigma_{\rho p}(W^2)$, such that, with $\sigma^{(\infty)}(W^2)$ replacing
$\sigma_{\rho p}(W^2)$, relation (\ref{2.42a}) is fulfilled at low energies.
At high energies,
$\sigma^{(\infty)}(W^2)$ is weakly dependent on energy, and it may even be
approximated by a constant of about 30 mb at 10\% accuracy. It is worth
noting that, within the limits of this approximation, the energy dependence
of photoproduction according to (\ref{2.31}) is entirely determined by the
generic two-gluon-exchange structure entering (\ref{2.31}) via
$I\left(m_0^2/\Lambda^2(W^2),m_0^2/\Lambda^2(W^2)\right)$. Since photoproduction
at high energies is well represented by both, (\ref{2.31}) and
(\ref{3.7}), the generic two-gluon-exchange structure and Pomeron exchange
are indeed seen to be dual representation of the same phenomenon.

\subsection{Comparing $\sigma_{\gamma^*p}(W^2,Q^2)$ in the GVD/CDP with
experiment for fixed $Q^2$ as a function of $W^2$.}

With $\sigma^{(\infty)}(W^2)\cong {\rm const}$ in the energy range relevant at HERA,
according to (\ref{2.9}), the energy dependence of the colour-dipole cross
section for fixed and sufficiently small interquark separation, $r_\perp$,
is determined by $\Lambda(W^2)$.

\begin{figure}[ht]
%\vspace*{-1.5cm}
\begin{center}
{\centerline{\epsfig{file=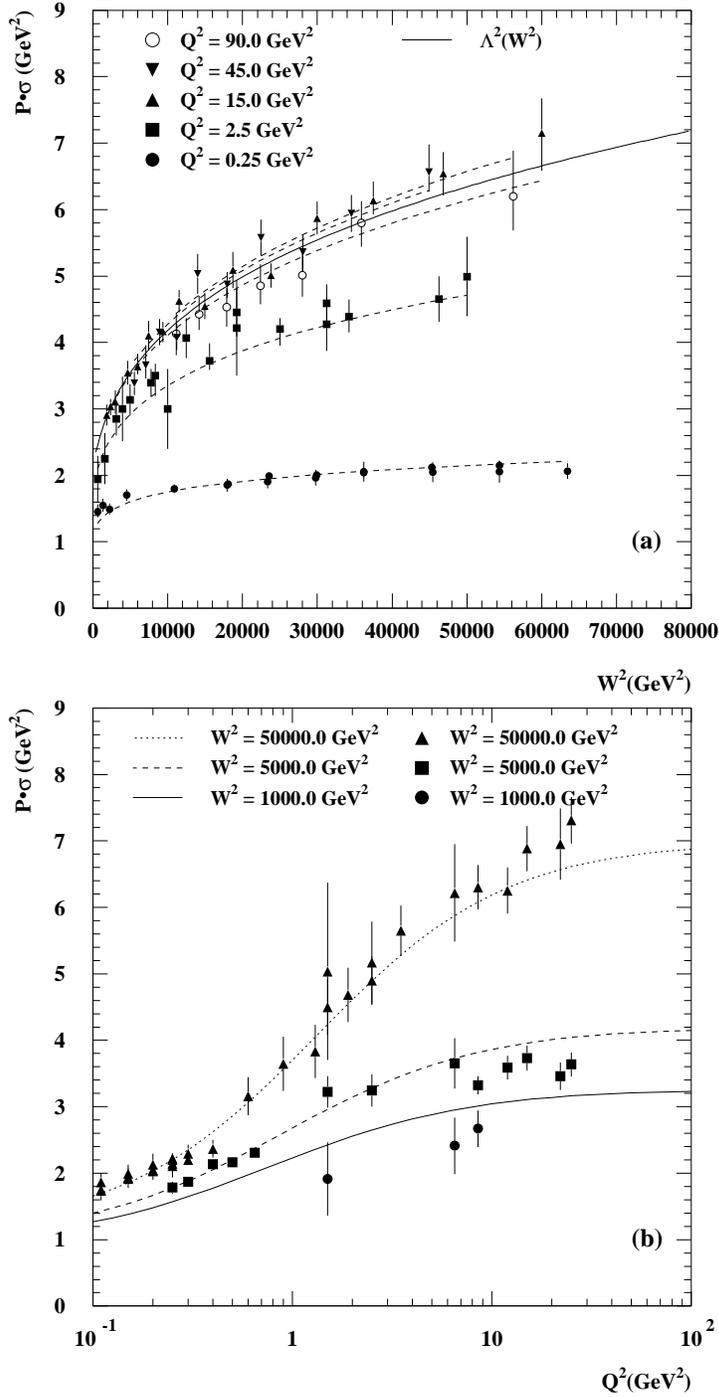,width=10.0cm}}}
\caption{
The quantity ${\rm P} \cdot \sigma_{\gamma^*
p}(W^2,Q^2)\equiv\frac{6\pi(q^2+m_0^2)}
{\alpha R_{e^+e^-}\sigma^{(\infty)}}\sigma_{\gamma^*p}(W^2,Q^2)$
defined in (\protect\ref{3.10a}) is shown as a function of $W^2$
at fixed $Q^2$ (a) and as a function of $Q^2$ at fixed $W^2$ (b).
The figure demonstrates that the data for $Q^2\gsim 10 {\rm GeV}^2$ yield
$\Lambda^2(W^2)$, where $\Lambda^2(W^2)$ determines the energy dependence of
the colour-dipole cross section at sufficiently small interquark separation
$r_{\perp}$.
\label{fig8i}}
\end{center}
\end{figure}

According to (\ref{2.23}), with (\ref{2.28}) and
(\ref{2.30}), the limiting behaviour of
$\sigma_{\gamma^*p}(W^2,Q^2)$, i.e.
\begin{equation}\label{3.10a}
\frac{6\pi}{\alpha
R_{e^+e^-}}(Q^2+m_0^2)\frac{\sigma_{\gamma^*p}(W^2,Q^2)}{\sigma^{(\infty)}}=
\left\{
\begin{array}{ll}
2(Q^2+m_0^2)\ln\frac{\Lambda^2(W^2)}{m_0^2},&\mbox{  for}Q^2\to 0\\
\Lambda^2(W^2),&\mbox{  for}Q^2\to \infty ,
\end{array}
\right.
\end{equation}
allows one to directly  deduce $\Lambda^2(W^2)$ from the experimental data
by plotting the left-hand side of (\ref{3.10a}) against $W^2$ at fixed
$Q^2$ or, alternatively, against $Q^2$ at fixed $W^2$ (and $x\leq
0.01$). Figure \ref{fig8i}
shows that the left-hand side of (\ref{3.10a}) approaches
$\Lambda^2(W^2)$ for $10 {\rm GeV}^2\lsim Q^2 \lsim
100 {\rm GeV}^2$. The upper limit on $Q^2$ corresponds to the upper limit
employed in fig \ref{fig5}a and used in the GVD/CDP fit to the data.
For fig.\ref{fig8i}, we use the value of $m_0^2=0.16 {\rm GeV}^2$ from
(\ref{3.9}), as well as $\sigma^{(\infty)}=80{\rm GeV}^{-2}\cong 31
\mbox{mb}$ according to fig.\ref{fig7}.

\begin{figure}[ht]
%\vspace*{-1.5cm}
\begin{center}
{\centerline{\epsfig{file=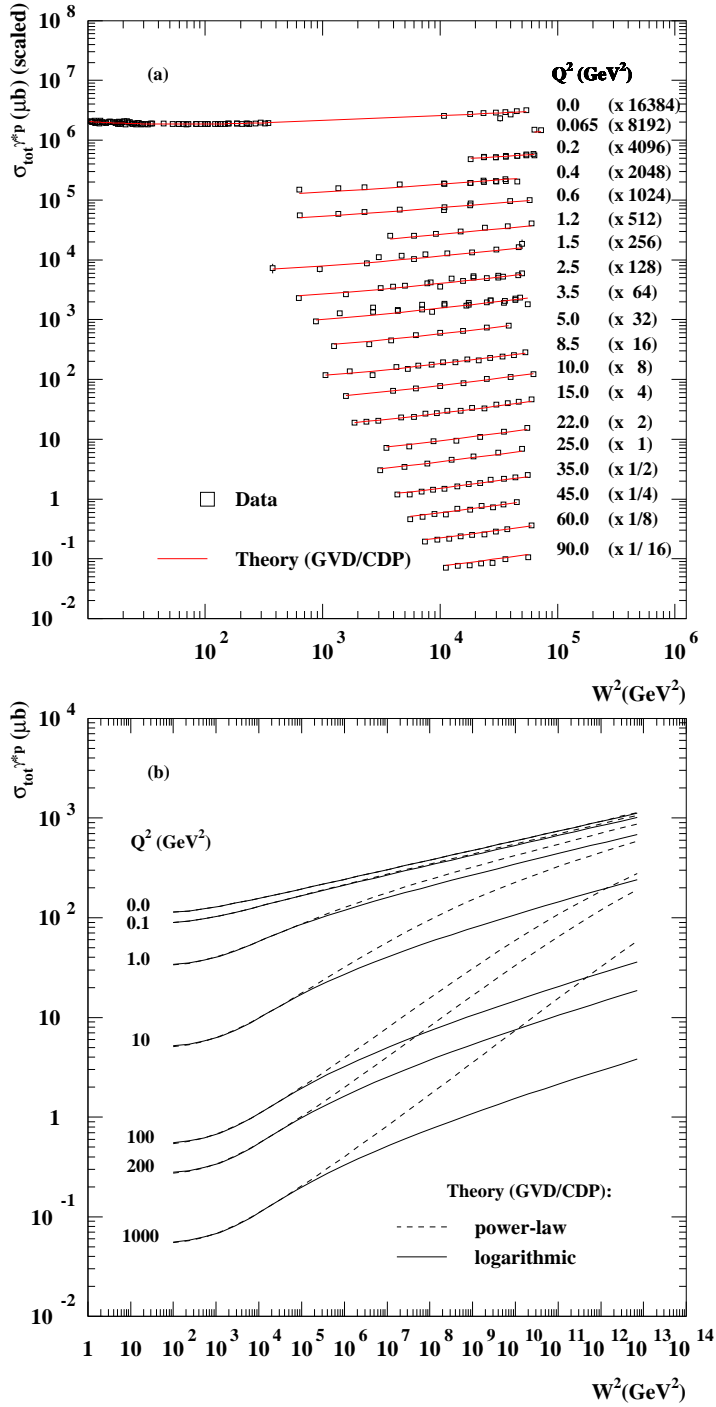,width=10.0cm}}}
\caption{The GVD/CDP predictions for $\sigma_{\gamma^* p}(W^2, Q^2)$ vs. $W^2$
at fixed $Q^2$\  a) in the presently accessible energy range
compared with experimental data for $x \le 0.01$,
\  b) for asymptotic energies.
\label{fig8}}
\end{center}
\end{figure}

Finally, in fig.\ref{fig8}a, we show the
GVD/CDP prediction in comparison with the
experimental data in the conventional
representation of $\sigma_{\gamma^*p}(W^2,Q^2)$
against $W^2$ for fixed $Q^2$. A subset of all data used in the
fits is presented for illustration.

The explicit analytical form of the theoretical expression for the cross
section, $\sigma_{\gamma^*p}(W^2,Q^2)$, allows us to investigate its
behaviour at energies far beyond the ones being explored at HERA. According
to (\ref{3.6}), with (\ref{2.28}) and (\ref{2.30}), at any sufficiently
large $Q^2\gsim\Lambda^2(W^2)$ (i.e. large $\eta$),
the cross section increases strongly with
energy, as $\Lambda^2(W^2)$, while finally, for sufficiently large
energy (i.e. small $\eta$),
the hadronlike dependence on energy of photoproduction, will be reached.
Explicitly this is demonstrated in fig.\ref{fig8}b. While the power-law
ansatz and the logarithmic one for $\Lambda^2(W^2)$ coincide at present
energies, they differ strongly in how asymptotics will be reached.
Unfortunately, the approach to the asymptotic behaviour is slow and can
hardly be verified experimentally in the foreseeable future, except,
possibly, by the energy dependence of precision data at small values of $Q^2\lsim
1\mbox{GeV}^2$.

\begin{figure}[ht]
\vspace*{-1.5cm}
\begin{center}
{\centerline{\epsfig{file=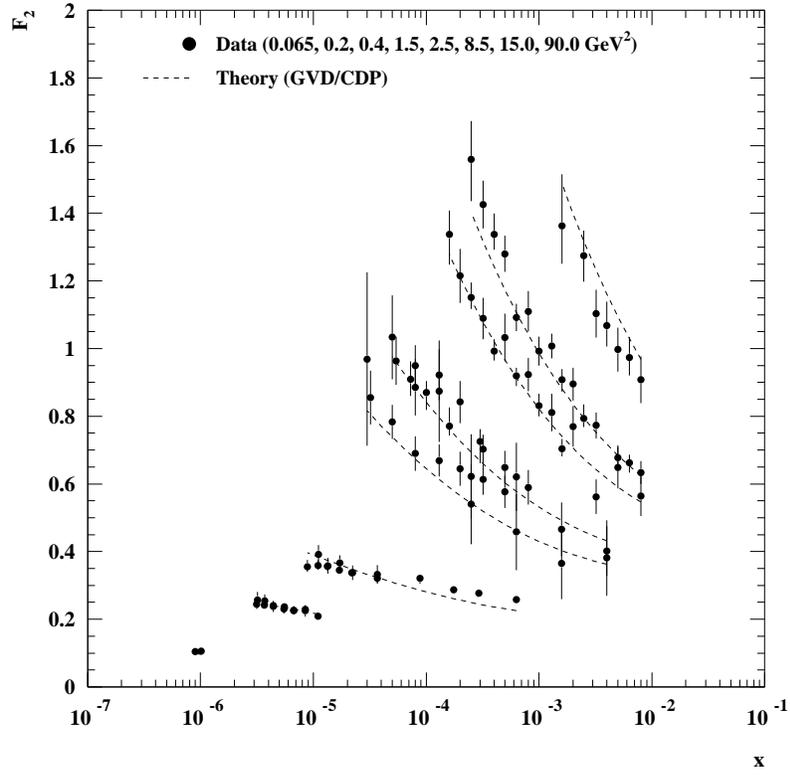,width=11.0cm}}}
\caption{The GVD/CDP prediction for the structure function $F_2(x,Q^2)$ in
comparision with the experimental data.\label{fig9}}
\end{center}
\end{figure}

Intuitively, a representation of the experimental data on DIS in the low-$x$
diffraction regime in terms of the virtual-photon-proton cross section seems
most appropriate and, in particular, reveals the scaling in $\eta$.
Nevertheless, for completeness, in fig.\ref{fig9}, we show
the data for the structure function $F_2(x,Q^2)$ together with the
theoretical results of the GVD/CDP.

\subsection{A reference to  related work}
The closest in spirit to the present investigation is the work by
Forshaw, Kerley and Shaw \cite{Forshaw}
and by Golec-Biernat and W\"usthoff \cite{Golec}.
While we agree with the general picture of low-x DIS drawn by these
authors, there are numerous essential differences though.
In our treatment, the dependence of the colour-dipole cross section on the
configuration variable $z$ is taken into account in contrast to
refs.\cite{Forshaw} and \cite{Golec}.
Our dipole cross section does not depend on $Q^2$, in agreement with the
mass-dispersion relations (\ref{2.13}) to (\ref{2.16}), but in distinction from the $Q^2$
(or rather $x$) dependence in ref.\cite{Golec}.
Decent high-energy behaviour at any $Q^2$ (``saturation'') follows from
the underlying assumptions
of colour transparency (the generic two-gluon-exchange structure) and
hadronic unitarity in distinction from the two-pomeron ansatz in
ref.\cite{Forshaw}
and in ref.\cite{dola}
that needs modification at energies beyond the ones explored at HERA\footnote{
For additional references and a report on a recent discussion meeting on the
CDP, we refer to ref.\cite{McDermott}}.

\section{Conclusion}
\setcounter{equation}{0}
In conclusion, a unique picture, the GVD/CDP, emerges for DIS in the
low-x diffraction region. In terms of the (virtual) Compton-forward-scattering
amplitude, the photon virtually dissociates into $(q \bar q)$ vector states
that propagate and undergo diffraction scattering from the proton as
conjectured in GVD a long time ago. Our knowledge on the
photon-$(q \bar q)$ transition from $e^+ e^-$ annihilation together with the
gluon-exchange dynamics from QCD allows for a much more detailed
theoretical description of $\sigma_{\gamma^* p}(W^2, Q^2)$ than available at
the time when the GVD approach was formulated.
In terms of the GVD/CDP, experiments on
DIS at low x measure the energy dependence of the
$(q \bar q)$/colour-dipole-proton
cross section, $\sigma_{(q \bar q)p} (r^2_\bot , z, W^2)$.
A strong energy dependence of
this cross section for small interquark separation
(not entirely unexpected within the GVD/CDP)
is extracted from the data at large $Q^2$.
The combination of colour transparency (generic two-gluon-exchange structure)
with hadronic unitarity then implies that for any interquark separation
the strong increase of the colour-dipole cross section with energy,
at sufficiently high energy, will settle
down to the smooth increase of purely hadronic
interactions.
The experimental data establish scaling in $\eta$ of $\sigma_{\gamma^* p}$.
As a consequence, at any fixed value of $Q^2$ (at low $x$),
$\sigma_{\gamma^* p}$ will eventually, at sufficiently high energy, reach
the hadronlike behaviour of photoproduction.

%\begin{center}
%\bf Acknowledgements
%\end{center}

\section*{Acknowledgements}
%\noindent
One of us (D.S.) thanks the theory group of the Max-Planck-Institut f\"ur
Physik in M\"unchen, where part of this work was done, for warm hospitality.
Thanks to Wolfgang Ochs for useful discussions, and
particular thanks to Leo Stodolsky for his insistence that there should
exist a simple scaling behaviour in DIS at low x.

\section*{Appendix}

\appendix

\renewcommand{\theequation} {\Alph{section}.\arabic{equation}}

\section{ Appendix A}
\setcounter{equation}{0}
In this Appendix we describe the steps leading from
expression (\ref{2.5}) to (\ref{2.13})--(\ref{2.16}).
The Fourier transform of the photon wavefunction as given,
for example, in Ref.~\cite{Cvetic}, is in the limit of
massless quarks
\begin{eqnarray}
{\cal M}_L^{(\lambda,\lambda^{\prime})}({\vec k}_{\perp}, z; Q^2)
& = &
- \frac{e_q z(1\!-\!z)}{(z(1\!-\!z) Q^2\!+\!k_{\perp}^2)}
\sqrt{Q^2} 2 \delta_{\lambda,-\lambda^{\prime}} \ ,
\label{ML}
\\
{\cal M}_{T, \pm 1}^
{(\lambda,\lambda^{\prime})}({\vec k}_{\perp}, z; Q^2)
& = &
- \frac{e_q}{(z(1\!-\!z) Q^2\!+\!k_{\perp}^2) \sqrt{2}}
k_{\perp} {\rm e}^{\pm {\rm i} \varphi} (2 z\!-\!1\pm\lambda)
\delta_{\lambda,-\lambda^{\prime}} \ ,
\label{MT}
\end{eqnarray}
where $e_q$ is the electric charge of quark $q$,
$\varphi$ is the azimuthal angle of ${\vec k}_{\perp}$ in the
plane perpendicular to the proton--photon axis of motion,
$\lambda$ and $\lambda^{\prime}$ denote twice the helicities
of the quarks $q$ and ${\bar q}$, and the signs $\pm$
correspond to the two transverse helicity state polarizations
$\epsilon^{\mu}(\pm) = (0,1/\sqrt{2}, \pm {\rm i}/\sqrt{2}, 0)$
of the massive photon in its rest frame. To obtain the total
cross section in (\ref{2.5}), the sum over the
helicities $\lambda$ and $\lambda^{\prime}$ is taken,
and in the case of the transversely (T) polarized photon $\gamma^{\ast}$
the average over the polarizations $P = \pm 1$
\begin{eqnarray}
\langle | {\cal M}_L ({\vec k}_{\perp},z; Q^2) |^2 \rangle
& \equiv & \sum_{\lambda, {\lambda}^{\prime} = \pm 1}
| {\cal M}_L^{(\lambda, {\lambda}^{\prime})}
({\vec k}_{\perp},z; Q^2) |^2
= \frac{ e_q^2 8 Q^2 z^2 (1\!-\!z)^2 }
{ \left[ z(1\!-\!z) Q^2 + {\vec k}^2_{\perp} \right]^2 } \ ;
\label{ML2sum}
\end{eqnarray}
\begin{equation}
\langle
{\cal M}_L ({\vec k}_{\perp}\!+\!{\vec l}_{\perp},z; Q^2)^{\ast}
{\cal M}_L ({\vec k}_{\perp},z; Q^2)
\rangle
 =  \frac{ e_q^2 8 Q^2 z^2 (1\!-\!z)^2 }
{ \left[ z(1\!-\!z) Q^2 + {\vec k}^2_{\perp} \right]
\left[ z(1\!-\!z) Q^2 + ({\vec k}_{\perp}\!+\!{\vec l}_{\perp})^2
\right] } \ ,
\label{MLMLsum}
\end{equation}
\begin{equation}
\langle | {\cal M}_T ({\vec k}_{\perp},z; Q^2) |^2 \rangle
 \equiv  \frac{1}{2}\!\sum_{P=\pm 1}
\sum_{\lambda, {\lambda}^{\prime} = \pm 1}
| {\cal M}_{T,P}^{(\lambda, {\lambda}^{\prime})}({\vec k}_{\perp},z; Q^2) |^2
= 2 e_q^2
\frac{ \left[ {\vec k}^2_{\perp} \left( z^2\!+\!(1\!-\!z)^2 \right)
\right] }
{ \left[ z(1\!-\!z) Q^2 + {\vec k}^2_{\perp} \right]^2 }
\ ,
\label{MT2sum}
\end{equation}
\begin{equation}
\langle
{\cal M}_T({\vec k}_{\perp}\!+\!{\vec l}_{\perp},z; Q^2)^{\ast}
{\cal M}_T({\vec k}_{\perp},z; Q^2)
\rangle
 =  2 e_q^2
\frac{ \left[
{\vec k}_{\perp}\!\cdot\!({\vec k}_{\perp}\!+\!{\vec l}_{\perp})
\left( z^2\!+\!(1\!-\!z)^2 \right) \right] }
{ \left[ z(1\!-\!z) Q^2\!+\!{\vec k}^2_{\perp} \right]
\left[ z(1\!-\!z) Q^2\!+\!({\vec k}_{\perp}\!+\!{\vec l}_{\perp})^2
 \right] } \ .
\label{MTMTsum}
\end{equation}
In (\ref{2.5}), the integration over ${\vec k}^{\prime}_{\perp}$
can be done trivially, resulting in
\begin{eqnarray}
\sigma_{\gamma^*_{T,L}p} (W^2,Q^2)  &= &
 {N_{\rm c} \over {16 \pi^3}}
\int dz \int d^2 l_\perp \tilde \sigma_{(q \bar q)p}
(\vec l^{~2}_\perp, z, W^2)
\nonumber\\
& \times &
{\Bigg \{}
\int_{\vert \vec k_\perp \vert \geq k_{\perp 0}}
d^2 k_\perp  \langle | {\cal M}_{T,L}
(\vec k_\perp, z; Q^2) |^2 \rangle
\nonumber \\
&
- &\int_{\vert \vec k_\perp \vert \geq k_{\perp 0},
\vert {\vec k}_{\perp}\!+\!{\vec l}_{\perp} \vert \geq k_{\perp 0} }
d^2 k_\perp
\langle {\cal M}_{T,L} ({\vec k}_{\perp}\!+\!{\vec l}_{\perp}, z; Q^2)^{\ast}
{\cal M}_{T,L}({\vec k}_{\perp}, z; Q^2) \rangle
{\Bigg \}} \ .
\label{sigtot2}
\end{eqnarray}
The multiple integrations in the above expression can be
rewritten in the following way. First we rename in all
the previous expressions the transfer momentum
${\vec l}_{\perp}$ to ${\vec l}^{\prime}_{\perp}$.
Then we rewrite in the above integrals the Fourier transform
${\tilde \sigma}_{(q{\bar q})p}(l^{\prime 2}_{\perp}, z, W^2)$
of the colour--dipole cross section as
\begin{equation}
{\tilde \sigma}_{(q{\bar q})p}(l^{\prime 2}_{\perp}, z) =
\int_0^{\infty} d l^2_{\perp}
{\tilde \sigma}_{(q{\bar q})p}(l^2_{\perp}, z)
\delta(l^2_{\perp}\!-\!l^{\prime 2}_{\perp}) \ .
\label{deltains}
\end{equation}
The integration over $d^2 l^{\prime}_{\perp}$ is then to be carried
out first. We denote the angle between ${\vec k}_{\perp}$ and
${\vec k}_{\perp}\!+\!{\vec l}^{\prime}_{\perp}$ as $\phi$.
We replace $d^2 l^{\prime}_{\perp}$ by
$d^2 k^{\prime}_{\perp}$ where we identify
${\vec k}^{\prime}_{\perp} \equiv {\vec l}^{\prime}_{\perp}\!+\!
{\vec k}_{\perp}$
\begin{eqnarray}
\lefteqn{
\int_0^1 dz \int d^2 l^{\prime}_{\perp}
{\tilde \sigma}_{(q {\bar q})p}(l^{\prime 2}_{\perp},z)
\int d^2 k_{\perp} f(k_{\perp},
| {\vec k}_{\perp}\!+\!{\vec l}^{\prime}_{\perp} |, \phi, z) }
\nonumber\\
& = &
\int _0^1 dz \int dl^2_{\perp} {\tilde \sigma}_{(q {\bar q})p}(l^2_{\perp},z)
\int d^2 k_{\perp}  \int d^2 k^{\prime}_{\perp}
\delta \left(
l^{2}_{\perp}\!-\!
({\vec k}^{\prime}_{\perp}\!-\!{\vec k}_{\perp})^2
\right)
f(k_{\perp}, k^{\prime}_{\perp},\phi, z) \ .
\label{mint1}
\end{eqnarray}
We now replace $d^2 k^{\prime}_{\perp}\!\equiv\!(1/2) d k^{\prime 2}_{\perp}
d \varphi_{k^{\prime}}$ by $(1/2) d k^{\prime 2}_{\perp} d \phi$, because
$\phi = \varphi_{k^{\prime}}\!-\!\varphi_{k}$. The subsequent integration
over $d \varphi_{k}$ gives $2 \pi$. The above expression reduces to
\begin{equation}
\frac{\pi}{2} \int_0^1 dz \int d l^2_{\perp}
{\tilde \sigma}_{(q {\bar q})p}(l^2_{\perp},z)
\int d k^2_{\perp} \int d k^{\prime 2}_{\perp} \int_0^{2 \pi} d \phi \;
\delta(2 k_{\perp} k^{\prime}_{\perp}\cos \phi\!-\!
k^2_{\perp}\!-\!k^{\prime 2}_{\perp}\!+\!l^2_{\perp})
f(k_{\perp}, k^{\prime}_{\perp}, \phi, z) \ .
\label{mint2}
\end{equation}
The integration over $d \phi$ can now be easily performed,
it fixes the value of $\cos \phi$ to a fixed value $\cos \Phi$, and
gives
\begin{equation}
\pi  \int_0^1 dz \int d l^2_{\perp}
{\tilde \sigma}_{(q {\bar q})p}(l^2_{\perp},z)
\int d k^2_{\perp}
\int_{(k_{\perp}\!-\!l_{\perp})^2}^{(k_{\perp}\!+\!l_{\perp})^2}
dk^{\prime 2}_{\perp}
{\tilde \omega}(k_{\perp}, k^{\prime}_{\perp}, l_{\perp})
f(k_{\perp}, k^{\prime}_{\perp}, \Phi, z) \ ,
\label{mint3}
\end{equation}
where
\begin{equation}
{\tilde \omega}(k_{\perp}, k^{\prime}_{\perp}, l_{\perp}) =
\frac{1}{2 k_{\perp} k^{\prime}_{\perp}}
\frac{1}{\sqrt{1 - \cos^2 \Phi}} \ , \quad
\cos \Phi =  \left(
\frac{ k^2_{\perp}\!+\!k^{\prime 2}_{\perp}\!-\!l^2_{\perp}}
{2 k_{\perp} k^{\prime}_{\perp}} \right) \ .
\label{cosphi}
\end{equation}
The integration limits for $k^{\prime 2}_{\perp}$
in (\ref{mint3}) are determined by the triangle
condition $\cos^2 \Phi \leq 1$. The fixed angle
$\phi = \Phi$ is the angle between the vectors
${\vec k}_{\perp}$ and ${\vec k}^{\prime}_{\perp}$
and, at the same time, the angle between the vectors
${\vec k}_{\perp}$ and ${\vec k}_{\perp}\!+\!{\vec l}_{\perp}$.

When we trade the variables $k^2_{\perp}$
and $k^{\prime 2}_{\perp}$ for $M^2\!=\!k^2_{\perp}/(z(1\!-\!z))$ (\ref{2.6})
and $M^{\prime 2}\!=\!k^{\prime 2}/(z(1\!-\!z))$ (\ref{2.7}), respectively,
taking into account the expressions (\ref{ML2sum})--(\ref{MTMTsum})
in (\ref{sigtot2}) and the transformed multiple integration form
(\ref{mint3}),\footnote{
We have to keep in mind that we replaced
${\vec k}_{\perp}\!+\!{\vec l}_{\perp}$
by ${\vec k}_{\perp}\!+\!{\vec l}^{\prime}_{\perp}\!\equiv\!
{\vec k}^{\prime}_{\perp}$
in the integrand expressions (\ref{ML2sum})--(\ref{MTMTsum})
and in (\ref{2.7}).}
we obtain for the transverse case
\begin{eqnarray}
\sigma_{\gamma^*_{T,p} (W^2,Q^2)}  &= &
 {N_{\rm c} \over {16 \pi^3}} \; 2 e_q^2 \; \pi \int_0^1 d z
\left[ z^2\!+\!(1\!-\!z)^2 \right]
 \int d l^2_{\perp} {\tilde \sigma}_{(q {\bar q})p}(l^2_{\perp},z, W^2)
\nonumber\\
&& \times {\Bigg \{}
\int_{m_0^2}^{\infty} d M^2
\int_{(M\!-\!L_{\perp}(z))^2}^{(M\!+\!L_{\perp}(z))^2} d M^{\prime 2}
\frac{M^2}{(Q^2\!+\!M^2)^2} \omega(M^2,M^{\prime 2}, L^2_{\perp}(z))
\nonumber\\
&& - \int_{m_0^2}^{\infty} d M^2
\int_{{\rm max} [ m_0^2, (M\!-\!L_{\perp}(z))^2]}^
{(M\!+\!L_{\perp}(z))^2} d M^{\prime 2}
\frac{(M^{2}\!+\!M^{\prime 2}\!-\!L^2_{\perp}(z))}
{2 (Q^2\!+\!M^2)(Q^2\!+\!M^{\prime 2})}
{\Bigg \}} \ ,
\label{sigT1}
\end{eqnarray}
where $L_{\perp}(z)\!\equiv\!l_{\perp}/\sqrt{z(1\!-\!z)}$, the
lower cutoff is $m_0^2\!\equiv\!k^2_{\perp 0}/(z(1\!-\!z))$,
and $\omega\!=\!z(1\!-\!z) {\tilde \omega}$
\begin{equation}
\omega(M^2,M^{\prime 2}, L^2_{\perp}(z)) = \frac{1}{2 M M^{\prime}}
\frac{1}{\sqrt{1 - \cos^2 \Phi}} \ , \quad
\cos \Phi = \frac{(M^{2}\!+\!M^{\prime 2}\!-\!L^2_{\perp}(z))}
{2 M M^{\prime}} \ .
\label{cosphi2}
\end{equation}
Using the ansatz (\ref{2.8}) for
${\tilde \sigma}_{(q {\bar q}) p}(l^2_{\perp}, z, W^2)$
allows for trivial integration over $l^2_{\perp}$,
resulting in the additional factor $\sigma^{({\infty})}/\pi$
and the replacement $L_{\perp}(z) \mapsto \Lambda(W^2)$.
Further, if we assume that $m_0^2$ is $z$--independent,
then the integration over $z$ can be done, giving  a factor $2/3$.
This then gives exactly the result
(\ref{2.11})--(\ref{2.13}) and (\ref{2.15}).
The square of the electric charge $e^2_q$ of the quark $q$
is replaced in general by the sum of the active
quark flavours $\sum e_i^2\!\equiv\!e_0^2\sum Q_i^2$ (\ref{2.12}),
and the number of quark colours is $N_{\rm c} = 3$.

Formulae (\ref{2.14}) and (\ref{2.16}) for the
longitudinal polarization can be
derived in a completely analogous way.

\section{ Appendix B}
\setcounter{equation}{0}

We restrict ourselves to giving the explicit expression for $I^{(1)}_T
\left( \eta, {{m^2_0} \over
{\Lambda^2(W^2)}}\right) $ and $I^{(1)}_L \left( \eta, {{m^2_0} \over
{\Lambda^2(W^2)}}\right)$. Evaluation of the integrals in (\ref{2.13}) and
(\ref{2.14}) yields
\begin{eqnarray}
\nonumber
\lefteqn{
   I^{(1)}_T
   \left( \eta,\mu \equiv {{m^2_0} \over {\Lambda^2(W^2)}}  \right)
}\\
%&&
                                        \label{b1}
= \frac{1}{2} \hspace*{-0.5cm}&&\ln
\frac{
        \eta-1+
        \sqrt{(1+\eta)^2-4\mu}
     }{
        2\eta
     }
 \\
%&&
\nonumber
+ \frac{
       1+2(\eta-\mu)
     }{
       2
       \sqrt{1+4(\eta-\mu)}
     }
\times \hspace*{-0.5cm}&&\ln
\frac{
       \eta
       \left(
             1+\sqrt{1+4(\eta-\mu)}
       \right)
     }{
       4\mu-1-3\eta+
       \sqrt{
           \Big(
               1+4(\eta-\mu)
           \Big)
           \Big(
               (1+\eta)^2-4\mu
           \Big)
       }
     }
\\
\nonumber
&&+\frac{\mu}{\eta}-1
\end{eqnarray}
and
\begin{eqnarray}
\nonumber
\lefteqn{
   I^{(1)}_L
   \left( \eta,\mu\equiv {{m^2_0} \over {\Lambda^2(W^2)}} \right)
}\\
%&&
=                                                    \label{b2}
 \frac{
       \eta-\mu
     }{
       \sqrt{1+4(\eta-\mu)}
     }
\times\hspace*{-0.5cm}&&
\ln
\frac{
       4\mu-1-3\eta+
       \sqrt{
           \Big(
               1+4(\eta-\mu)
           \Big)
           \Big(
               (1+\eta)^2-4\mu
           \Big)
       }
     }{
       \eta
       \left(
             1+\sqrt{1+4(\eta-\mu)}
       \right)
     }
\\
\nonumber
&+& \left(
    1-\frac{\mu}{\eta}
  \right).
\end{eqnarray}
One easily checks that
$I^{(1)}_L
   \left( \eta,\mu\equiv{{m^2_0} \over {\Lambda^2(W^2)}} \right) \to 0
$ for $\eta \to \mu$.

Summing $I^{(1)}_L$ and $I^{(1)}_T$ yields
(\ref{2.24}).

\newpage

\noindent

\newpage

\end{document}